\documentclass[10pt,letterpaper]{article}
\usepackage[top=0.85in,left=2.75in,footskip=0.75in]{geometry}

\usepackage{amsmath,amssymb}

\usepackage{changepage}

\usepackage[utf8x]{inputenc}

\usepackage{textcomp,marvosym}

\usepackage{cite}

\usepackage{nameref,hyperref}

\usepackage[right]{lineno}

\usepackage{microtype}
\DisableLigatures[f]{encoding = *, family = * }

\usepackage[table]{xcolor}

\usepackage{array}

\newcolumntype{+}{!{\vrule width 2pt}}

\newlength\savedwidth

\newcommand\thickhline{\noalign{\global\savedwidth\arrayrulewidth\global\arrayrulewidth 2pt}%
\hline
\noalign{\global\arrayrulewidth\savedwidth}}

\newcommand\sbullet[1][2]{\mathbin{\vcenter{\hbox{\scalebox{#1}{$\bullet$}}}}}

\raggedright
\usepackage[aboveskip=1pt,labelfont=bf,labelsep=period,justification=raggedright,singlelinecheck=off]{caption}

\makeatletter
\renewcommand{\@biblabel}[1]{\quad#1.}
\makeatother

\usepackage{lastpage,fancyhdr,graphicx}
\usepackage{epstopdf}
\usepackage{verbatim}
\fancyheadoffset[L]{2.25in}
\fancyfootoffset[L]{2.25in}
\usepackage{hyperref}
\hypersetup{colorlinks=true, citecolor=orange, urlcolor=blue, linkcolor=magenta}
\usepackage{makecell}
\usepackage{multirow}

\definecolor{color00}{rgb}{0.5450980392156862, 0.5254901960784314, 0.5098039215686274}
\definecolor{color01}{rgb}{0.5450980392156862, 0.0392156862745098, 0.3137254901960784}
\definecolor{color02}{rgb}{0.2823529411764706, 0.4627450980392157, 1.0}
\definecolor{color03}{rgb}{0.7333333333333333, 1.0, 1.0}
\definecolor{color04}{rgb}{0.5411764705882353, 0.16862745098039217, 0.8862745098039215}
\definecolor{color05}{rgb}{1.0, 0.5137254901960784, 0.9803921568627451}
\definecolor{color06}{rgb}{0.9333333333333333, 0.9098039215686274, 0.6666666666666666}
\definecolor{color07}{rgb}{1.0, 0.9607843137254902, 0.9333333333333333}
\definecolor{color08}{rgb}{1.0, 0.8431372549019608, 0.0}
\definecolor{color09}{rgb}{1.0, 0.27058823529411763, 0.0}
\definecolor{color10}{rgb}{0.803921568627451, 0.3568627450980392, 0.27058823529411763}
\definecolor{color11}{rgb}{1.0, 0.6274509803921569, 0.47843137254901963}
\definecolor{color12}{rgb}{0.9333333333333333, 0.23137254901960785, 0.23137254901960785}

\newcommand\ale[1]{{\color{black}#1}}
\newcommand\fab[1]{{\color{black}#1}}

\newcommand\alee[1]{{\color{black}#1}}

\begin{document}
\vspace*{0.2in}

\begin{flushleft}
{\Large
\textbf\newline{Firms' Challenges and Social Responsibilities during Covid-19: a Twitter Analysis}  
}
\newline
\\
Alessia Patuelli\textsuperscript{1,*},
Guido Caldarelli\textsuperscript{2,3},
Nicola Lattanzi\textsuperscript{1},
Fabio Saracco\textsuperscript{4}

\bigskip
\textbf{1} AXES - Laboratory for the Analysis of Complex Economic Systems, IMT School for Advanced Studies, Lucca, Italy
\\
\textbf{2} DSMN, Università Ca’ Foscari, Mestre (Ve), Italy
\\
\textbf{3} CNR- Institute of Complex Systems, Unit ``Sapienza", Rome, Italy
\\
\textbf{4} Networks Unit, IMT School for Advanced Studies, Lucca, Italy
\bigskip

%
%




* alessia.patuelli@imtlucca.it

\end{flushleft}
\section*{Abstract}

The Covid-19 pandemic caused disruptive effects for individuals, firms, and societies. In this paper, we offer insights on the major issues and challenges firms are facing in the Covid-19 pandemic, as well as their concerns for Corporate Social Responsibility (CSR) themes. \ale{To do so}, we investigate large Italian firms’ discussion on Twitter in the first nine months of the pandemic. Specifically, we ask:  How is firms’ Twitter discussion developing during the Covid-19 pandemic? Which CSR dimensions and topics do firms discuss?  To what extent do they resonate with the public?\\ 
\fab{We downloaded all Twitter posts from 1st of March, 2020, to 17th of November, 2020 by the accounts of the largest Italian firms, i.e. those with 250 or more employees. We then built the bipartite network of accounts and hashtags and, using an entropy-based null model as a benchmark, we projected the information contained in the network into the accounts layers, identifying a network of accounts, in which a link indicates a non trivial similarity in terms of their usage of hashtags. }\ale{We find that the conversation is focused around 13 communities, 10 of which include Covid-19 themes. The core of the network is formed of 5 communities, which deal with environmental sustainability, digital innovation and safety. 
Firms' ownership type does not seem to influence the conversation. 10 communities out of 13 mention hashtags related to CSR, with the environmental and social dimensions as the prevalent ones. Interestingly enough, the social dimension seems more relevant in the communities dealing with digital innovation and safety. However, the relevance of CSR hashtags is very small at the single message level, \fab{but }with some peculiarities arising in specific communities. Also, users engage very little on CSR themes, with some differences showing in a few communities.\\ 
\alee{Overall, our paper highlights the role of network methods on Twitter data as a tool which can support managers and policy makers to design their strategies and decision making, capturing firms’ emerging issues and relevant themes.} With the relevance of CSR dimensions and stakeholder engagement varying depending on the community, we also propose network methods as an effective way to capture the differences in firms' and stakeholders' CSR concerns. }


\section{Introduction}

The Covid-19 pandemic impacted multiple business dimensions and forced firms to rethink the way they do their businesses, coping with what has been called ``the new normal”~\cite{Ratten2020}.
While most research~\cite{Almeida2020,Amore2020,Crick2020a,Evans2020,Ivanov2020,Juergensen2020} explored Covid-19 effects on firms using conceptual, survey or interview-based methods, only one paper, to the best of our knowledge, examined firms’ challenges through social media, focusing on the effects on supply chains~\cite{sharma2020covid}. Among others, the analysis of online social networks is a promising tool to understand current issues, trends and challenges firms are facing~\cite{Pilar2019}. 

In the last 15 years, social media revolutionized communication, making it cheaper and faster than ever before and opening a new channel for firms to \fab{directly} communicate and interact with their stakeholders. 
An increasing number of firms opened social media pages~\cite{Reilly2014}, which also allowed a new research stream to develop \cite{li2019identifying,sharma2020covid}.
Management research focusing on social media mostly dealt with Corporate Social Responsibility (CSR) and stakeholder engagement, with manual collection~\cite{Reilly2014} and labelling~\cite{Manetti2016}, and linear regressions~\cite{Gomez-Carrasco2020,Yang2020,Yang2017}. 
These methods bring a number of limitations. First, they are able to capture only a small number of firms, which is generally less than a hundred~\cite{Saxton2019,Yang2020,Yang2017}. Second, they follow \emph{a priori} approaches~\cite{Saxton2019}, that do not allow new insights to emerge from data.  To the best of our knowledge, few papers examined firms’ communication strategies adopting network methods~\cite{Chae2018a,Colleoni2013}, and none of them analyzed firms’ discussion during the pandemic. As a data-driven approach could reveal complex patterns~\cite{choudhury2018machine}, we adopt a complex network analysis to understand firms’ conversation on Twitter at the beginning of the Covid-19 pandemic. Doing so, we aim at understanding firms’ issues and challenges during the Covid-19 crisis, as reflected in the Twitter discussion. 
\ale{Specifically, we aim at answering three explorative research questions: How is firms’ Twitter discussion developing during the Covid-19 pandemic? Which CSR dimensions and topics do firms discuss? To what extent do they resonate with the public? 

\fab{
In order to detect similarities in the usage of hashtags, we represent the system as a bipartite network between accounts and hashtags. We then project the information contained in the network on the layer of accounts, to find users using similar hashtags. The network is then validated via a comparison with a null model. In this sense, we opt for the Bipartite Configuration Model (\emph{BiCM},~\cite{Saracco2015}), i.e. an entropy-based networks null-model,  since it is unbiased by construction~\cite{Cimini2018,Squartini2017}. The validation procedure is the the one proposed in~\cite{Saracco2017}.\\}
First, we find that firms' discussion on Twitter form 13 communities \fab{of accounts}, 10 of which \fab{touch upon} Covid-19 themes. The core of the network is made of 5 communities, which focus on environmental sustainability, digital transformation, remote working, digitalization and safety. This highlights that firm's dialogue  on environmental, digital innovation and safety themes was central in Italian large firms' discussion on Twitter at the beginning of the Covid-19 pandemic. Second, 10 communities out of 13 use CSR hashtags. Among the communities, the environmental and social dimensions are prevalent, with the economic one being overlooked. Interestingly, communities dealing with digital innovation and safety focus more on the social dimension than the environmental one, which contrasts the main literature \cite{pedersen2010modelling} while confirming that CSR dissemination is context-dependent \cite{dahlsrud2008corporate}.  However, when we focus on the message level, results show that CSR messages are a minority, with some peculiarities appearing in some communities. Last, users seem to interact little on CSR topic, showing that stakeholder engagement on these themes on Twitter is still scarce.}

\ale{
Our paper brings methodological, theoretical and practical contributions.
From a methodological perspective, we integrate complex network analysis in management research, thus introducing a data-driven approach. Following a non-linear approach, with no a-priori hypotheses, themes and communities freely emerge from data, allowing similarities and differences between firms to arise in real time.}

\alee{This allows our contributions to span across different management fields, including the literature on firms and social media, CSR and stakeholder engagement.  
On the practical side, our research offers a tool for monitoring current challenges and issues firms are facing. This can be useful for policy makers and managers to orient their strategies and decision making \cite{ruhi2014social} and to understand how firms' and users' perceptions of CSR themes vary, reflecting different beliefs of what responsibilities firms have towards society~\cite{pedersen2010modelling}.

}

\section{Previous findings}

\subsection{Firms, social media and the Covid-19 pandemic}

In the last 20 years, Internet changed the ways and the speed with which firms communicate and relate to their stakeholders.  At first, the social dimension of website communication was focused on blogs, which allowed firms to communicate their “personality” and incorporate a relationship strategy based on connectivity and dialogue with users. Blogs started dialogic online relationships between firms and their public.  

When they first appeared around 15 years ago, social media were aimed at social networking with friends, family and colleagues~\cite{Saxton2014}.  They opened new paths and ways for firms to communicate with their stakeholders, shifting from a hierarchical one-to-many communication to a many-to-many collaborative type of communication~\cite{Colleoni2013}. 
Thus, social media became platforms for firms to share advertising, marketing and public relationship strategies~\cite{Saxton2014}.\par
Among social media, Twitter is popular for business communication purposes~\cite{Gomez-Carrasco2020}, and is the one with the highest adoption rates among large firms~\cite{Yang2017}. \textcolor{black}{It is increasingly used in academic research, with most contributions arising from the computer and information science, and communications, while developing also in business and economics \cite{bruns2014topology}. While Twitter research is mostly focused on the marketing \cite{jansen2009twitter} and  financial fields \cite{bollen2011twitter}, it is increasingly being used in the management and strategy fields, for example to monitor firms’ CSR dissemination and stakeholder engagement \cite{Lee2013}, the emerging trends in technology \cite{li2019identifying}  or to understand the issues and challenges firms are facing on specific themes \cite{sharma2020covid}. 
}

This literature mostly studied firms' communication on Twitter with traditional methodologies. On the data collection side, manual collection methods \cite{Reilly2014} prevail, while on the data analysis side, manual labelling \cite{Manetti2016}, and linear regressions \cite{Yang2020,Gomez-Carrasco2020,Yang2017} are the majority. However, some studies are experimenting with new approaches based on network methods, with a prevailing focus to understand CSR communication and stakeholder engagement on social media \cite{Colleoni2013,Pilar2019}.

There is a prolific new research line which focuses on the impacts of Covid-19 on firms and their reactions (among others:~\cite{Eggers2020,Juergensen2020,Amore2020,Bartik2020,Rapaccini2020}) and there is an extended research about the Covid-19 as perceived on Twitter (see, for instance,~\cite{Rovetta2020, Celestini2020, Cinelli2020, Gallotti2020, yang2020covid19,Caldarelli2020b,lopreite2021early}), but, surprisingly,  researchers just started to explore firms’ conversation on Covid-19 on Twitter. To the best of our knowledge,  only one paper~\cite{sharma2020covid} has studied firms’ discussion during Covid-19, focusing on the pandemic’s effects on supply chains. 
However, the pandemic impacted multiple business dimensions, including firms’ financial resources and business models~\cite{Amore2020}, with first reactions ranging from increasing servitization, digitalization~\cite{Juergensen2020, Evans2020, Almeida2020}, and cooperation~\cite{Crick2020}. Managers and entrepreneurs had to rethink the way they do their businesses, learning to face a totally unexpected crisis~\cite{Wankmuller2020} and to cope with a “new normal” \cite{Ratten2020}.
To the best of our knowledge, there is no paper currently addressing how firms’ conversation on social media is evolving during Covid-19, thus providing insights on the current issues and challenges firms are facing~\cite{Pilar2019}.

Thus, we want to investigate how firms’ Twitter discussion is developing during the Covid-19 pandemic.

\subsection{Firms, CSR dissemination and stakeholder engagement}

\par
\textcolor{black}{Although the concept of CSR has no universal meaning and has been widely defined~\cite{carroll1999corporate}, it generally refers to the relationships and responsibilities a business has towards society, interpreted as the communities of stakeholders a firm interacts with~\cite{snider2003corporate,dahlsrud2008corporate,carroll1999corporate}, going beyond what is required by law~\cite{mcwilliams2001corporate}. What CSR concerns in practical terms depends on the historical and cultural context a firm is settled in, and it can also reflect the issues a firm is facing in a particular period~\cite{snider2003corporate}, which it have been changing and increasing fast after the start of globalization~\cite{dahlsrud2008corporate}.
}
\par
\textcolor{black}{
While the advent of the Internet pushed firms to disseminate their social reporting online~\cite{snider2003corporate}, social media opened a new channel for firms to disseminate their actions and interact with their stakeholders~\cite{Gomez-Carrasco2020}.
}

\par
Previous studies show that firms use social media to disseminate their social, environmental or sustainability orientation and results, but it seems that social media are not fully exploited as a mean to interact with stakeholders~\cite{Manetti2016}. One explanation is legitimacy theory~\cite{guthrie1989corporate}, which argues that companies are on social media to show their presence in the digital world rather than to report their results or engaging their stakeholders. \textcolor{black}{ In this paragraph we will discuss previous research findings, focusing on (i) how the CSR dimensions are concerned in firms' discussion on Twitter and (ii) to what extent stakeholders (here considered as Twitter users) interact with firms. 
}

Most research finds that firms' social media posts about CSR topics are a minority. 
Considering together firms' Facebook posts, Tweets and Youtube videos in a 3-months period in 2014,~\cite{Manetti2016} show that CSR posts are only  7\% of all posts, thus suggesting that firms do not use Twittter to disseminate their CSR activities. Focusing on the top 50 companies from the Fortune list of 2010,~\cite{Gomez2016} find similar results, highlighting that firms mostly post about topics non-related to CSR (among others, posts include marketing purposes, promotion of products and services), with only a small percentage of posts regarding CSR. This is also consistent with~\cite{Etter2013}, who finds that in general business accounts 14.5\% of tweets are about CSR topics, while they rise in CSR-dedicated Twitter pages, with more than 70\% tweets on average about CSR issues. However, there are some conflicting results. For example, taking Spanish banks after the euro zone crisis,  CSR appears as a material topic discussed on Twitter~\cite{Gomez-Carrasco2020}.  

Few papers use network (visualization) methods, thus showing which communities around CSR topics arise. For example, selecting firms with a Twitter account dedicated to CSR and using network and machine learning methods,~\cite{Colleoni2013} finds that firms and users create several independent communities around CSR themes, instead of a unique and connected network. 
Combining structural topic modeling and network methods,~\cite{Chae2018} explore how the Twitter discussion around the hashtag \#CSR evolves,  resulting in 20 different topics (the most prevalent one being company strategy, followed by community charity, CSR teams, business ethics, …), most of them related to one or more topics, and seven without any significant correlation with others. The paper concludes that Twitter is used as a mean to share about multiple CSR dimensions, some of them are related and they change over time.  
Following a similar approach, starting from the  hashtag \#sustainability, \cite{Pilar2019} adopt social networks methods, finding top themes associated with \#sustainability (namely, Innovation, Environment, Climate Change, Corporate Social Responsibility, Technology, and Energy) and 6 communities (Environmental Sustainability, Sustainability Awareness,  Renewable Energy and Climate Change, Innovative Technology, Green Architecture, and Food Sustainability). 
\newline

Overall, existing research seem to agree that CSR communication on Twitter is a minority and results are quite fragmented. However, there are mixed findings and data-driven methods have been rarely used. Also following~\cite{Etter2013} call for a more in-depth study on the topics firms address on Twitter when communicating on CSR, we want to investigate which CSR dimensions and topics firms discuss.

While the aforementioned results relate to one-way communication on CSR on Twitter, a few papers also question the extent to which users interact with firms on Twitter, as a measure of stakeholder engagement.
\textcolor{black}{
Being a separate but related concept to CSR, stakeholder engagement can be interpreted under many different theoretical perspectives and has been variously defined. Following \cite{greenwood2007stakeholder}, “stakeholder engagement is understood as practices the organisation undertakes to involve stakeholders in a positive manner in organisational activities” (p. 315).
Social media are increasingly used as a tool to measure stakeholder engagement, with the assumption that social media users are part of the firm’s stakeholders \cite{bonson2013set}.
Although some preliminary evidence shows that Twitter is less used as a mean to engage stakeholders compared to Facebook  \cite{manetti2017stakeholder}, it is still considered an ideal tool for firms to engage with their stakeholders, with measures like retweeting activity that can be interpreted as an implicit endorsement of the content of the message  \cite{Saxton2019}.} 
Papers exploring firms' use of Twitter for engaging their stakeholders use different approaches and findings are quite mixed. 
In general, it seems that social media  are not used in their full capacity to interact and engage with stakeholders on CSR topics. This is what both \cite{Gomez2016} and \cite{Manetti2016} found, the former highlighting that firms interact very little on CSR topics, the latter finding that firms' posts concerning stakeholder engagement are only the 0.22\% of total messages.
This is somewhat consistent with \cite{Etter2013}, who finds that non-CSR  tweets have a higher interactivity than CSR ones. However, \cite{Saxton2019} find opposite results, with CSR messages gaining a greater audience reaction than non-CSR messages, in particular when certain CSR topics (e.g. environment or education) are discussed and when the post uses a hashtag related to the topic. However, popularity was measured as a binary variable (retweeted message vs not retweeted message), not taking into account the extent to which a message is retweeted. Similarly, focusing on specific CSR dimensions, a few papers study how stakeholder engagement varies depending on the type of CSR communication. For example, using the very broad categorization of core CSR (information directly linked to the firm's core business) versus supplementary CSR  (information about actions detached from the firm's core business), \cite{Gomez-Carrasco2020} find that different categories of stakeholders react differently to firm's CSR tweets depending on the content that is shared. However, results do not allow to further distinguish which CSR dimensions resonate most. 
\par
As the literature shows very mixed results in stakeholder engagement on CSR topics on Twitter, we want to investigate to what extent the CSR dimensions resonate with the public. 

\section{Materials and methods}

\subsection{Data set}
We selected Italy for our study as it is one of the countries first and more severely hit by the pandemic~\cite{Amore2020}. 
Although  large firms are a minority in European businesses~\cite{EuropeanCommissionReportSme2019}, they usually have more stakeholders than small and medium sized firms, are considered to have a bigger impact on society~\cite{Campopiano2015} and higher resources to invest in reporting and accountability~\cite{giacomini2020environmental} . Therefore, we expect them to have a higher adoption rate of social media.  
We relied on AIDA\footnote{AIDA is a database containing financial and commercial historical data from approximately 540,000 firms operating in Italy. It relies on official data retrieved from the Italian Registry of Companies and the Italian Chambers of Commerce. AIDA contains data at the firm level, which include information about the firm’s characteristics (e.g. location, industry, the year it was founded), the ownership and governance structure (e.g. the name of each shareholder and board member, the respective ownership share) and financial data (balance sheet, profit and loss accounts data, and ratios).} (\href{ https://aida.bvdinfo.com/}{ https://aida.bvdinfo.com/}), the Italian section of Bureau Van Dijk (\href{https://www.bvdinfo.com/}{https://www.bvdinfo.com/}), to select firms. 

We selected all Italian active firms from with 250 or more employees, resulting in 3.870 firms. 
We downloaded data about firms' characteristics (e.g. their names, location, ownership type, ATECO code\footnote{ATECO is the classification of economic activities used by the Italian Institute of Statistics (ISTAT). It is the national translation of NACE, which is the classification of economic activities in the European Union. It stands for “Statistical classification of economic activities in the European Community” and is derived from the French “Nomenclature statistique des activités économiques dans la Communauté européenne”. The latest classification is NACE Rev. 2, which was implemented from 2007.} and website) and financial data (e.g. total assets and revenues). As at the time of download (April 2020) not every firm's financial data was updated with information on 2019, the financial data we downloaded refer to 2018. 
We manually searched the firms’ websites as indicated in AIDA database, checking in each website if a Twitter account was available. For each website, we searched the home page and “contacts” sections. 
We found that 936 companies out of 3870 (24\%) displayed a Twitter account in their website. A few of them indicated multiple Twitter accounts on their websites. 

\subsubsection{Data collection and first cleaning}
To collect all messages written by the selected accounts, we made use of the \href{https://developer.twitter.com/en/docs/twitter-api/v1/tweets/timelines/overview}{GET statuses/user\_timeline} Twitter API. Out of the original 936 accounts, 6 accounts were not active in the period examined, while 6 were not found (probably they were deactivated). The final data set was composed by 1.6 M of tweets.
\newline

The data set was cleaned further: we focused on firms writing tweets in the period from the 1st of March, 2020, to 17th of November, 2020, which includes the beginning of the Covid-19 pandemic. 
In order to focus on the proper set, we selected firms that wrote a message before and after the 1st of March, 2020, to explore their activity over a longer period. Literally, the \href{https://developer.twitter.com/en/docs/twitter-api/v1/tweets/timelines/overview}{GET statuses/user\_timeline} API permits to download the last circa 3200 messages written by the account\footnote{For more details, visit \href{https://developer.twitter.com/en/docs/twitter-api/tweets/timelines/introduction}{Twitter developer webpage}.}: if the account is extremely active, these 3200 will cover just an extremely limited period and thus cannot be considered for the definition of a narrative over the entire period under analysis.
It is the case, for instance, of the accounts of publishing houses owning newspapers: the Twitter accounts were used as the accounts of the newspapers, thus for communicating the latest piece of news. Other examples include public transport accounts (messages referred to traffic alerts), internet and mobile service companies (those accounts were either used to advertise the latest promotions or for the customer service) or football teams (commenting the results of the match). Selecting  accounts which tweeted before and after the 1st of March 2020 also permits to get rid of those accounts that were active in the past, but did not contribute to the discussion during the pandemic.
The resulting firms in our data set are a total of 417 different active accounts and 917864 different messages.
\ale{Even though our research is mostly based on publicly available online data (i.e. Twitter messages), we believe it is ethically appropriate not to disclose the firms' names, usernames and quotes \cite{stommel2021ethical}.}
\\ 

\subsubsection{Hashtag cleaning}
For each firm in the data set, we extracted the hashtag used, in order to study similarities in the communication strategies. Only 401 accounts out of 417 (i.e. the 96\%) used at least one hashtag in the period under study.\\
We find frequent mistakes in typing the hashtags: to avoid considering as different two hashtag referring to the same subject, we implemented the Edit distance, as implemented by the \texttt{py\_stringmatching} python module~\cite{py_stringmatching} (more details in the hashtag data cleaning can be found in the appendix~\ref{app:hashtags}). After this cleaning, 11475 different hashtags resulted in the data set.

\fab{\subsubsection{CSR hashtags}}
\textcolor{black}{To recognise hashtags related to CSR issues,  the widely known classification of sustainability proposed by~\cite{elkington1997cannibals} was used. It distinguishes between three dimensions: (i) the environmental dimension (i.e. attention of the company towards environmental issues), (ii) the social dimension (i.e. the relationship between business and society – consumers, employees and stakeholders in general), (iii) the economic dimension (i.e. socio-economic or financial aspects, including describing CSR in terms of business operations). In order to select relevant keywords, we adapted and integrated previous approaches. As ~\cite{Gamerschlag2011}  distinguish the environmental and social dimension, using keywords from the GRI\footnote{The Global Reporting Initiative (GRI) is is an independent, international organization aimed at developing voluntary reporting guidelines for CSR. The GRI Standards are among the world’s most widely used standards for CSR reporting. For more information, please visit: \href{https://www.globalreporting.org/}{ https://www.globalreporting.org/} } (Global Reporting Initiative) 2010 standard, we updated this list with the latest GRI 2019 standards (Italian translation), also including the economic dimension. To create a comprehensive list of CSR keywords, we also integrated keywords from existing research~\cite{Etter2013,Manetti2016}  and added them to their relevant CSR area. 
To do this, the first and fourth authors discussed the category for each item, and placed it in the category they both agreed on. 
}
\newline
\subsection{The bipartite semantic network and its validated projection}\label{ssec:bicm}
We represented the system as a bipartite network, i.e. a network in which the nodes are divided into two disjointed classes (called \emph{layers}) and edges are not allowed between nodes of the same group. One layer represents the different accounts, while the second layer represents the various hashtags. A hashtag and an account are connected if, in the period analysed, the given account used the selected hashtag at least once.
\newline

We use the strategy described in~\cite{Saracco2017} to project the information contained in the original system on the layer of the accounts and detect similarly communicating firms. In a nutshell, this approach consists of 3 main steps: first, we consider the network of the co-occurrences of links, as they are observed in the real system. In our case, since we are interested in projecting the information contained in our system onto the company layer, for each couple of users, the co-occurrences are the number of hashtags used by both accounts.
Second, we define a proper null-model as a benchmark. In the present application, we use a maximum entropy null-model, i.e. a model maximally random, but for the information contained in some constraints~\cite{Cimini2018,Squartini2017}. In particular, we can discount the information contained in the degree sequence, as in the Bipartite configuration Model (\emph{BiCM},~\cite{Saracco2015}; more details on the null-model can be found in the appendix~\ref{app:bicm}). In this way, in our analysis we are considering the attitude of the various accounts in using hashtags (that is, the degree of the various accounts) and the number of users using the given hashtag (i.e. the degree of the different hashtags).
Finally, we compare the observed co-occurrences with the expectations of the null-model: if those are statistically significant, i.e. much higher than the ones predicted by the model and such that the disagreement cannot be explained by the constraint of the  null-model, they are then validated. We then put an edge in the projection connecting the two accounts whose co-occurrences are statistically significant. 
The result of the projection is a monopartite undirected network of accounts in which a connection indicates non trivial similarities in terms of the usage of hashtags. The entire approach is described in more detail in the appendix~\ref{app:inferring}.
\newline

For the implementation of the BiCM, we used the recently released python module {\tt\href{https://pypi.org/project/NEMtropy/}{NEMtropy}}, presented in~\cite{vallarano2021fast}.

\section{Results and Discussion}

\subsection{General statistics}
\begin{figure}[htb!]
\includegraphics[width=.5\linewidth]{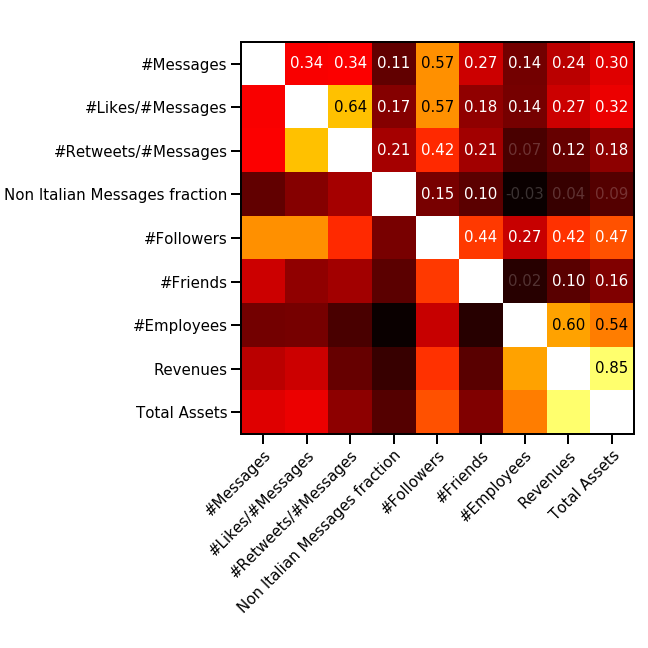}
\includegraphics[width=.5\linewidth]{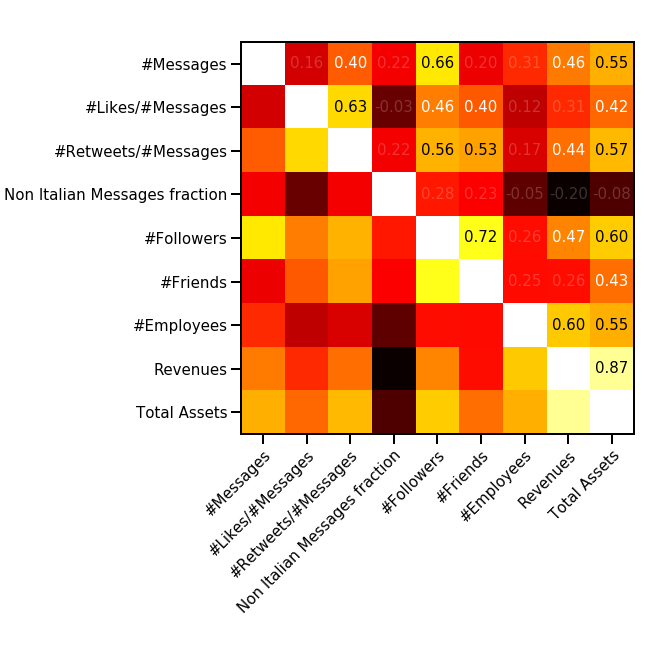}
\caption{\fab{{\bf Spearman correlations among different \ale{financial and online} data.} On the left panel the correlation matrix for the entire data set: all correlations between online and financial quantities are weak. This changes once we focus on specific ATECO codes, due to their peculiarities: on the right panel the same correlation matrix for sector 62 (\emph{Computer programming, consultancy and related activitie}). A relatively strong correlation is present between the number of followers and the total assets, for instance. This behaviour may be due to the importance of the communications in online social networks for this sector, which increases with a firm's resources.}}
\label{fig:correlation}
\end{figure}

Before considering the accounts that have been validated by the projection techniques described in the section above, we first present some features of the data set, that includes \ale{financial data} from AIDA and \ale{online data} from Twitter.\\  
\fab{Over the entire set of firms, the correlations among financial and online data remain limited: in Fig.~\ref{fig:correlation} we show the related Spearman correlation\footnote{Due to the non Gaussian distribution of the various quantities, Pearson correlation is not justified.} matrix. The correlations between quantities related to the companies' activity on online social networks and those regarding their financial performance are in general weak and rarely significant. Actually, even among Twitter quantities, the correlations are quite weak, with a few exceptions: the number of likes per message and the number of retweets per message (0.64), the number of followers and the number of messages (0.57), and the number of followers and the number of likes per message (0.57). \\ 

Interestingly enough, once we focus on a single ATECO code, as in the right panel of Fig.~\ref{fig:correlation} for code 62 (\emph{Computer programming, consultancy and related activities}), the situation appears slightly different. There are significant and relatively strong correlations between, for instance, the number of followers and the total assets (0.60), and the total number of messages and the total assets (0.55). This difference is probably due to the specific sector: being related to information technologies, the appearance of these companies in online social networks is an important part of their marketing strategies and depends on a firm's size (as measured in the total assets).}

Focusing on Twitter data, the retweeting activity is always the most frequent one, for instance representing nearly the 80\% of the actions on the platform during the Covid-19 Italian debate~\cite{Caldarelli2020a}. Interestingly enough, probably due to the different role of firms' Twitter accounts, the retweeting activity in the present data set is extremely limited: just the 22.3\% of the messages collected are retweets, while the others are original messages. \fab{A similar behaviour was detected for verified users~\cite{Caldarelli2020a}: the authors interpreted these findings maintaining that verified users are the main drivers for the development of a debate. Here, analogously, firms participate to the discussion by introducing new arguments, even if the accounts are not necessarily verified: the frequency of verified accounts in our data set is quite limited, close to 17\%.}

The most frequently used hashtag in the dataset is “covid”, which appears 4106 times. “Coronavirus” is the third most used hashtag, appearing 2120 times. The other most frequently used hashtags are mostly related to public utilities themes and digitalization.\\ 
\fab{We remind that we just selected firms' accounts, not keywords, to download the messages we analysed. Nevertheless, almost all accounts used hashtags related to pandemic, due to the obvious impact that it had on every activity in the period.}

\subsection{Validated network of Twitter accounts}

\begin{figure}[htb!]
\includegraphics[width=.75\linewidth]{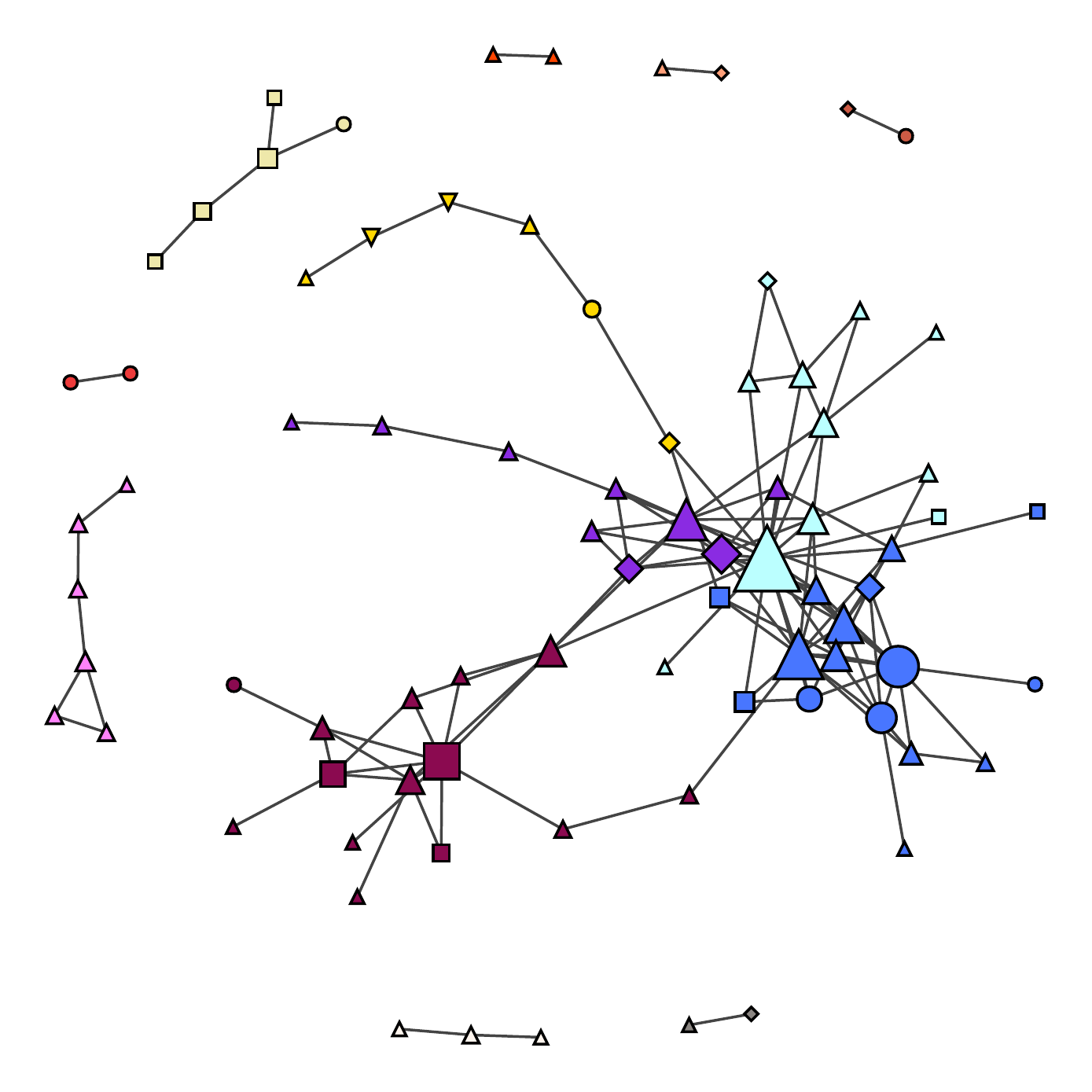}
\caption{{\bf Validated projection of the network of firms.} The network is composed by 80 firms and 135 links. The dimension of each node is proportional to its degree, i.e. the number of connections. The various colors represent the different communities. The shape of the nodes indicate the GUO (Global Ultimate Owner): rectangles are firms owned by the state or other public bodies, diamonds are firms owned by mutual \& pension funds / nominal / trust funds, circles are firms owned by individuals or families (family firms), while triangles are companies; the rest are diamond shaped.}
\label{fig:network}
\end{figure}

\begin{figure}[htb!]
\includegraphics[width=.5\linewidth]{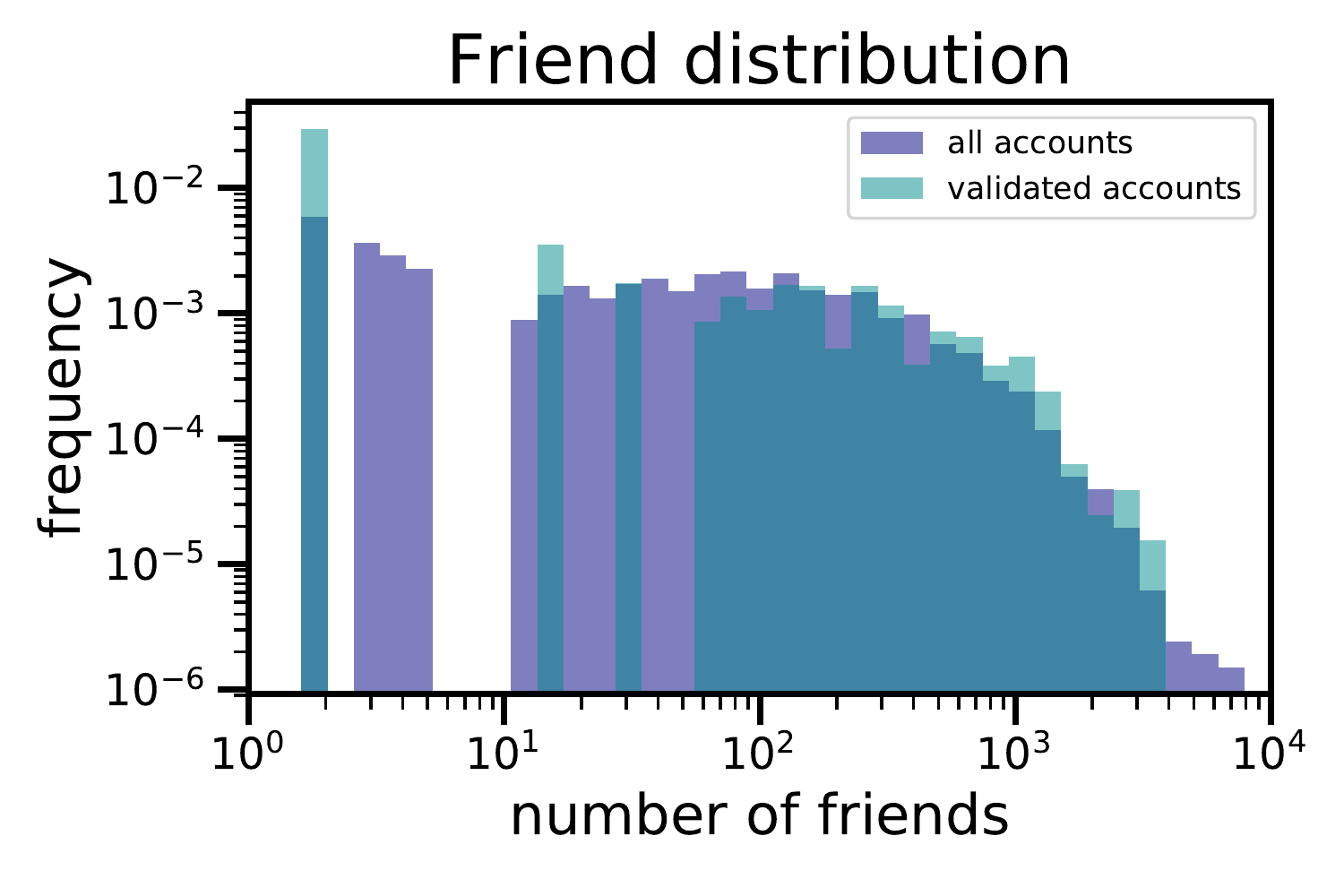}
\includegraphics[width=.5\linewidth]{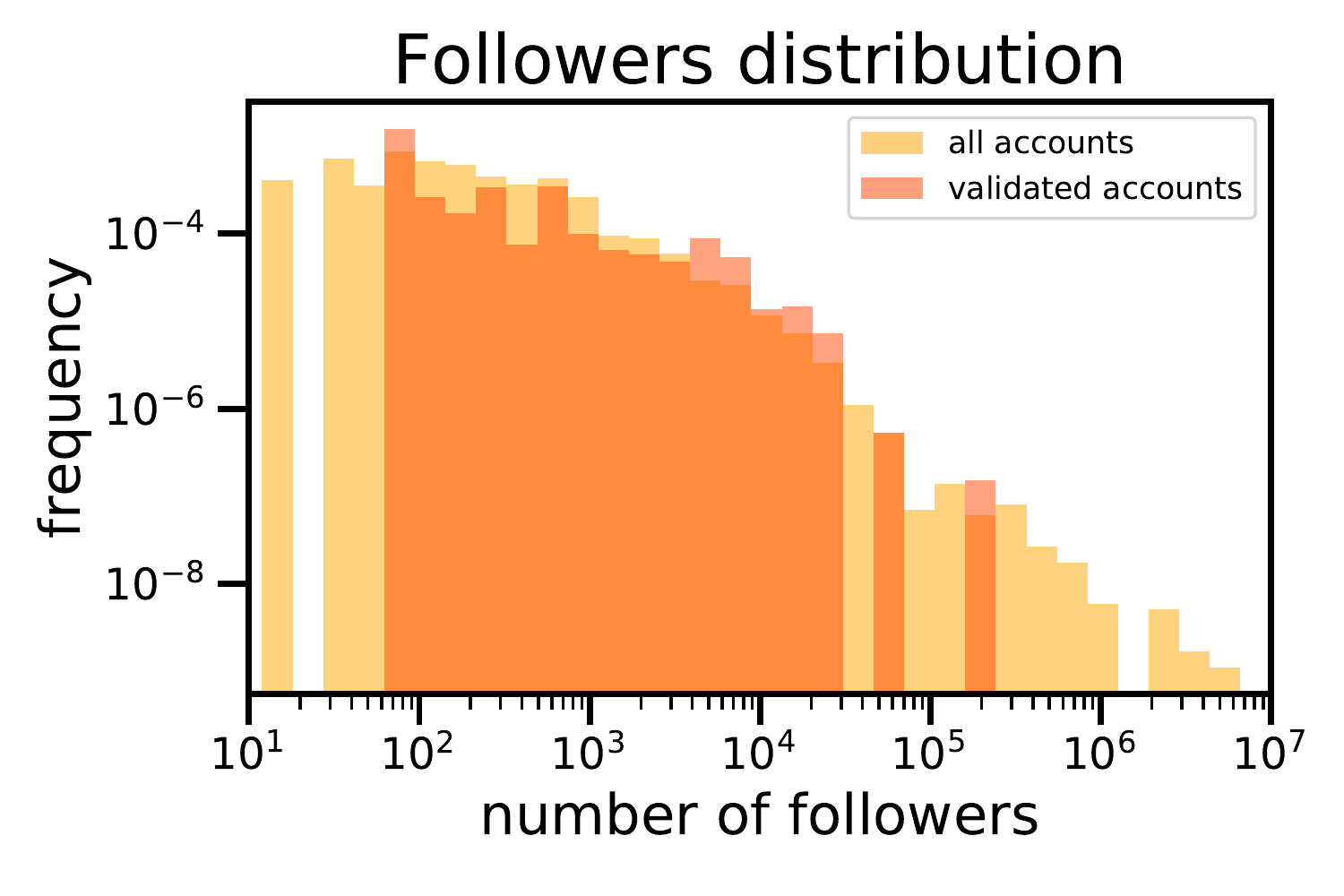}
\includegraphics[width=.5\linewidth]{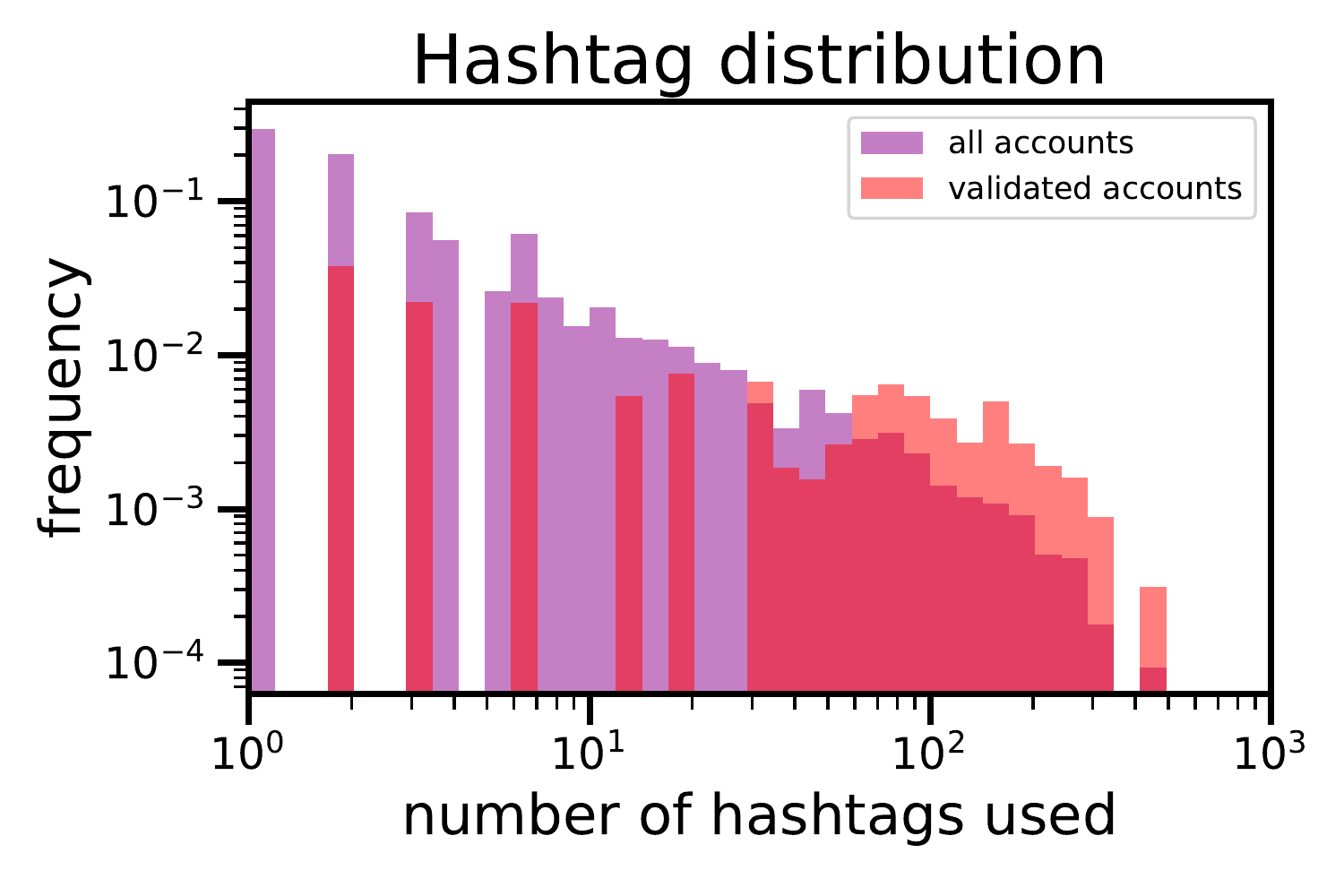}
\includegraphics[width=.5\linewidth]{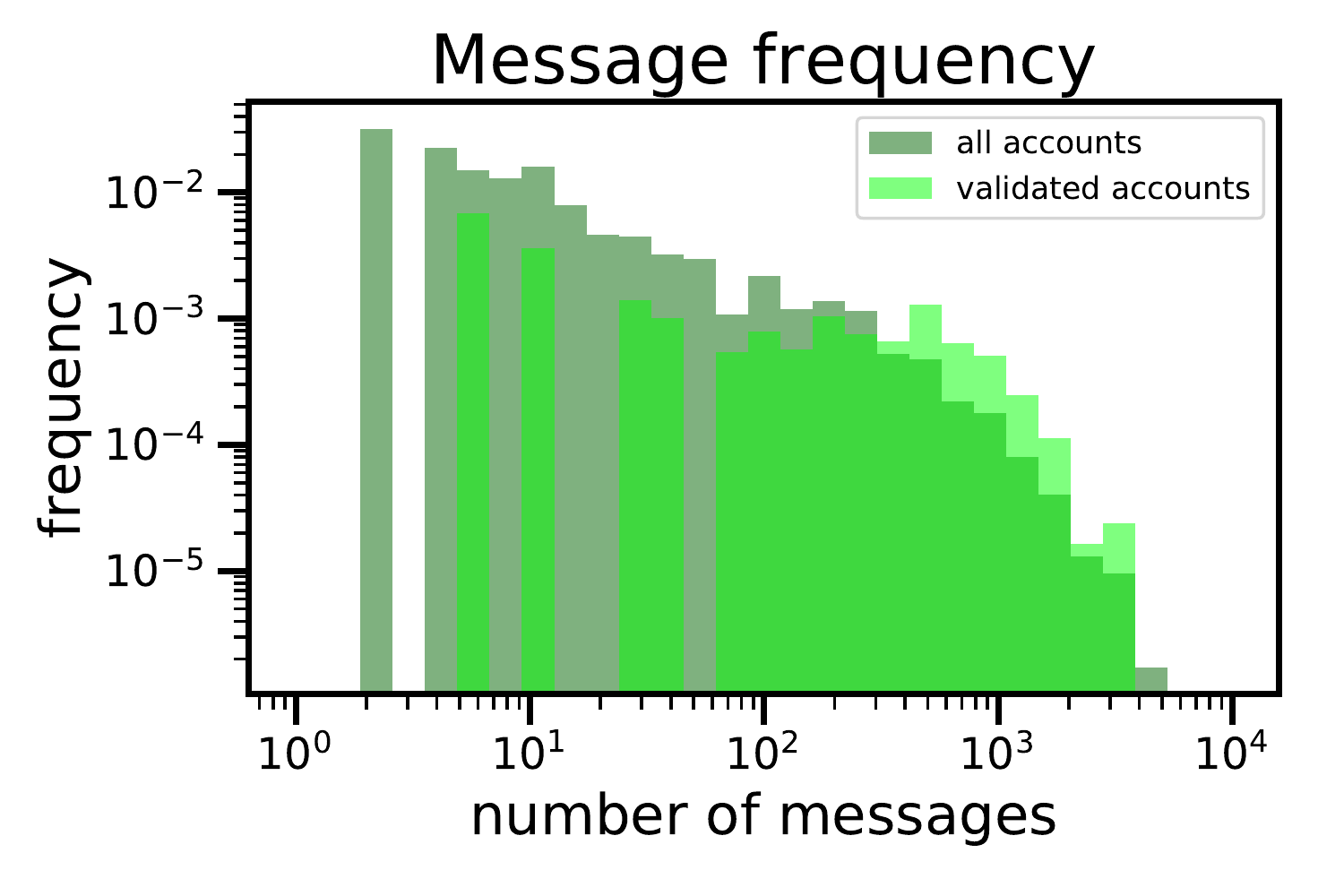}
\caption{\fab{{\bf Properties of nodes in the validated network vs. the ones in the entire set.} The various distributions show that the validated accounts are those that have a greater number of friends (top left panel), use  greater number of hashtags (bottom left panel) and write more messages (bottom right panel). Interestingly enough, the validated nodes are not the most popular, i.e. those with the highest number of followers (top right panel). In order to check the most popular accounts, we focused on accounts with more than $10^6$ followers. 
In fact, the number of their messages is extremely limited, while their use of hashtags is extremely focused on their activities. This may be  related to their strategies, to remark the exclusiveness of their products. More details can be found in the main text.}}
\label{fig:validated_network_statistics}
\end{figure}

\begin{figure}[htb!]
\includegraphics[width=.5\linewidth]{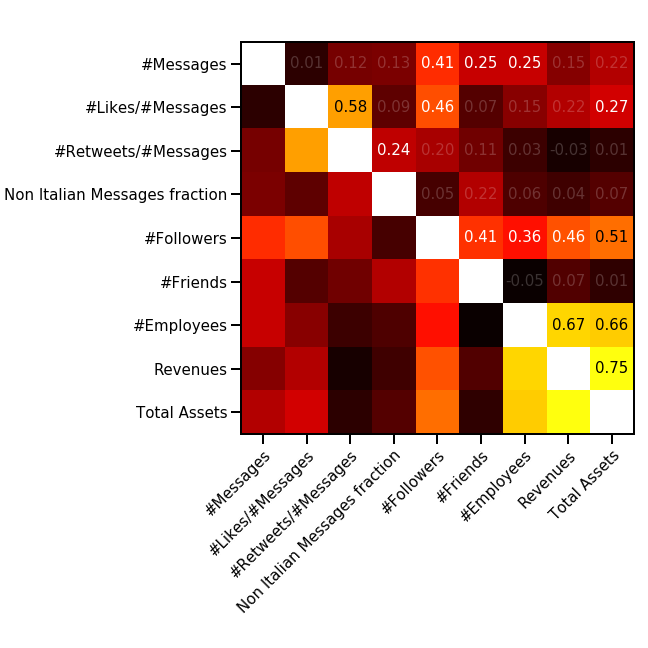}
\includegraphics[width=.5\linewidth]{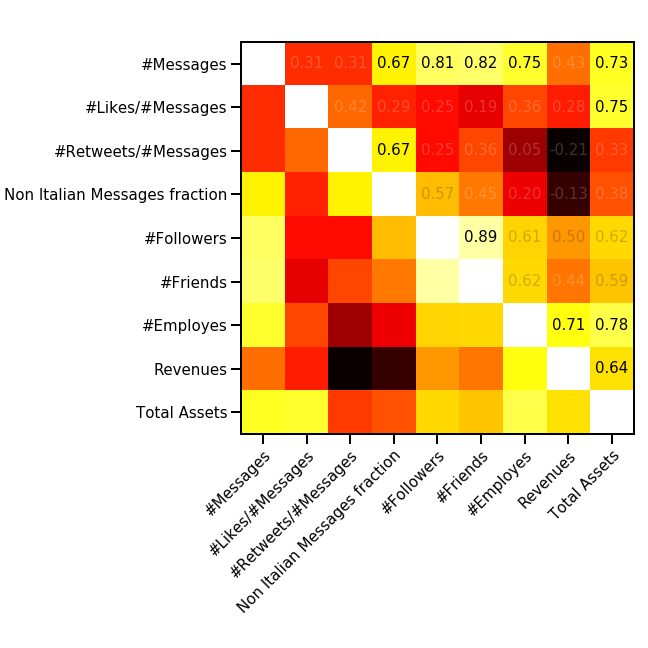}
\caption{\fab{{\bf Spearman correlations among different financial and online data for firms in the validated network.} On the left panel the correlation matrix for the validated network: all correlations between online and financial quantities are weak, as in the entire data set (see left panel of Fig.~\ref{fig:correlation}), but for the one between the number of followers and the total assets (0.51). Again, the situation changes once we focus on a specific ATECO code and it is even more striking than for the entire data set (see left panel of Fig.~\ref{fig:correlation}). In fact, the same correlation matrix for sector 62 (\emph{Computer programming, consultancy and related activitie}) shows a strong correlation, for instance, between the total number of messages and the total assets (0.73) and between the number of Likes per message and the total assets (0.75). In fact, the validated accounts are those that follow a common strategy in the usage of hashtags and the importance of their appearance online increases relatively to the dimension of the company.}}
\label{fig:correlation_validated}
\end{figure}

\fab{The result of the procedure described in Sec.~\ref{ssec:bicm} is a monopartite network of Twitter accounts composed by 80 firms and 135 links, in which links indicate a non trivial similarity in the usage of hashtags  (see Fig.~\ref{fig:network}). Interestingly enough, while being heavy users of messages and hashtags and having a great number of friends, the validated accounts are not among the most popular ones, see Fig.~\ref{fig:validated_network_statistics}. Indeed, the most popular users, i.e. those with more than a million followers, are luxury brands; those accounts wrote many messages, but they limit the usage of hashtags to their merchandise, probably in order to focus on the exclusiveness of their products. In this sense, let us remark that the validated nodes are those that contribute to the formation of common narratives, shared among various accounts; these extremely popular users, instead, are marking their originality and do not intervene substantially to shape common discussions.\\ 
Even in the case of the firms in the validated network, there are limited correlations between online and financial data, see the left panel of Fig.~\ref{fig:correlation_validated}: the only relatively strong (and significant) one is between the number of followers and the total assets (0.51). Nevertheless, when we focus on a specific ATECO code, correlations can be much stronger and mix the online data from Twitter with the financial ones: in the right panel of Fig.~\ref{fig:correlation_validated} we observe quite a strong correlation between the number of messages and the total assets (0.73) or between the number of likes per message and the total assets (0.75). A similar behaviour was observed in the analysis of the entire data set, see Fig.~\ref{fig:correlation}; the correlations on the 62 ATECO code, were, nevertheless, a little weaker than the ones observed here.}


\fab{By running the Louvain algorithm~\cite{Blondel2008}, we identified 13 communities of accounts, identified by nodes of different colors in Fig.~\ref{fig:network}. These groups show a certain homogeneity in terms of companies' sectors (for more details, see Tables~\ref{tab:ATECO_0} and \ref{tab:ATECO_1} in the appendix). In particular, the greatest communities in the core of the network mostly include computer programming, consultancy and related activities ({\color{color02} $\bullet$}), electricity, gas, steam and
air conditioning supply ({\color{color01} $\bullet$}), employment ({\color{color03} $\bullet$}), activities of head offices and management consultancy ({\color{color04} $\bullet$}) and human health ({\color{color08} $\bullet$}). 
Instead, }\ale{the ownership type (in term of the \emph{Global Ultimate Owner}, or GUO\footnote{As defined by Bureau Van Dijk, the GUO (Global Ultimate Owner) is the individual or entity at the top of the corporate ownership structure, thus indicating the highest parent company.}) does not seem to influence the conversation\fab{: the GUO is represented in Fig.~\ref{fig:network} as different shapes of the nodes and no significant pattern can be detected.} This is surprising, as some kinds of firms communicate in a different way, especially about CSR and environmental issues. For example, family firms are found to communicate on CSR differently from non-family firms~\cite{Campopiano2015} and publicly owned firms disclose less on environmental topics (contrary to the expectations~\cite{argento2019sustainability,giacomini2020environmental}).} 
Although ownership does not seem to influence the conversation, it is worth noting that the dialogue on environmental sustainability is mostly carried out by firms active in the public utilities sector.

\subsubsection{The usage of hashtags in the validated network: firms' discussion during the pandemic}
\fab{Remarkably, among the 13 communities detected, }10 of them have a Covid-related theme in their main hashtags (e.g. ``coronavirus”, ``covid”, ``iorestoacasa”, the latter one meaning ``stayhome”), meaning that firms’ discussion is strongly related to the pandemic. 
\ale{The core of the network is made of 5 interconnected communities. Based on their most frequently used hashtags, we named them ``Digital Transformation" ({\color{color02}$\sbullet$}), ``Remote Working" ({\color{color03}$\sbullet$}), ``Digitalization" ({\color{color04}$\sbullet$}), ``Environmental Sustainability" ({\color{color01}$\sbullet$}) and ``Safety" ({\color{color08}$\sbullet$}), which emerged as the main themes that firms discussed at the beginning of the Covid-19 pandemic.} The names assigned to all communities can be found in Table~\ref{tab:recap}.\\ 

\begin{table}[htb!]
    \centering
    \begin{tabular}{+ll+} 
 \thickhline
 \textbf{Comm.} & \textbf{Theme} \\ 
 \thickhline
 {\color{color02}$\sbullet$} & Digital Transformation \\
 {\color{color01}$\sbullet$} & Environmental Sustainability \\
 {\color{color03}$\sbullet$} & Remote Working \\
{\color{color04}$\sbullet$} & Digitalization \\
{\color{color05}$\sbullet$} & Restart\\
{\color{color08}$\sbullet$} & Safety \\
{\color{color06}$\sbullet$} & Mobility\\
{\color{color07}$\sbullet$} & Slowing the Spread\\
{\color{color00}$\sbullet$} & Solidarity\\
{\color{color09}$\sbullet$} & Space\\
{\color{color10}$\sbullet$} & Nursing Homes\\
{\color{color11}$\sbullet$} & Locality\\
{\color{color12} $\sbullet$} & Reading\\
\thickhline
\end{tabular}
\caption{{\bf Identification between themes and the community displayed in Fig.~\ref{fig:network}:} as it can be further observed, the themes the different groups of accounts deal with are closely related to their ATECO codes, i.e. to their sector.}
    \label{tab:recap}
\end{table}

\begin{table}[htb!]
    \hskip-1.5cm
    \begin{tabular}{+c|c|c+c|c|c+}
    \thickhline
        \textbf{Comm.} & \textbf{Hashtag} & \textbf{Occurrence} & \textbf{Comm.} & \textbf{Hashtag} & \textbf{Occurrence}\\
\thickhline
          {\color{color02}  $\sbullet$} & smartworking & 15 & {\color{color05}  $\sbullet$} & ripartiamodallitalia & 5 \\
          \cline{2-3} \cline{5-6} & cloud & 15  && covid & 5  \\ 
          \cline{2-3} \cline{5-6}& covid & 13 &&  italia & 5  \\ 
          \cline{2-3} \cline{5-6}& innovazione & 13 & & coronavirus & 5 \\ 
          \cline{2-3} \cline{5-6}& ai & 12 & & aprile & 4 \\ 
          \cline{2-3} \cline{5-6}& sicurezza & 12 & & turismo & 4 \\ 
          \cline{1-3} \cline{5-6}
          {\color{color01}  $\sbullet$} & sostenibilità & 13 & & estate & 4 \\ 
          \cline{2-3} \cline{5-6} & covid & 12 & & estatepostcovid & 4 \\ 
          \cline{2-3} \cline{5-6} & innovazione & 11 & & sostenibilità & 4   \\
          \cline{2-3} \cline{5-6} & greendeal & 11 & & coop & 4 \\ 
          \cline{2-3} \cline{4-6} & energia & 11 &  {\color{color08}  $\sbullet$} & covid & 6 \\ 
          \cline{1-3} \cline{5-6}
          {\color{color03}  $\sbullet$} & covid & 11 & & italia & 6  \\ 
          \cline{2-3} \cline{5-6}& digitale & 10 &  & coronavirus & 6  \\ 
          \cline{2-3} \cline{5-6}& coronavirus & 10  & & diabete & 5 \\ 
          \cline{2-3} \cline{5-6}& lavoro & 9 & & scuola & 5 \\ 
          \cline{2-3} \cline{5-6}& smartworking & 9 &  & sanità & 5\\ 
          \cline{1-3} \cline{5-6}
          {\color{color04}  $\sbullet$} & smartworking & 9 &  & mascherine & 5 \\ 
          \cline{2-3} \cline{4-6}& covid & 9 &  {\color{color06}  $\sbullet$} & settembre & 5 \\ 
          \cline{2-3} \cline{5-6}& lavoro & 8 &  & coronavirus & 5 \\  
          \cline{2-3} \cline{5-6}& webinar & 7 & & agosto & 5 \\  
          \cline{2-3} \cline{5-6}& coronavirus & 6 &  & covid & 5 \\  
          \cline{2-3} \cline{5-6}& digitale & 6 &  & ottobre & 5 \\  
          \cline{2-3} \cline{5-6}& economia & 6 & & maggio & 5 \\
          \cline{2-3} \cline{5-6}&&&& marzo & 5 \\  
          \cline{1-3} \cline{5-6}
          \thickhline 
    \end{tabular}
    \caption{{\bf The top 5 frequent hashtags for the communities in the Largest Connected Component of Fig.~\ref{fig:network}}. Due to the great number of \emph{ex aequo}, the top 5 most frequent hashtags is, in general, longer than 5. }
    \label{tab:hashtag_0}
\end{table}

In the “Environmental Sustainability” community ({\color{color01}$\sbullet$}), the most important hashtags are ``sostenibilità” (meaning ``sustainability”), ``covid”, ``innovazione” (``innovation”), ``greendeal”, ``energia” (``energy”), thus showing that the environmental themes are relevant and linked to the innovation ones, see Table~\ref{tab:hashtag_0}. This community is mostly formed by firms managing national infrastructures, and public utilities. 
Three communities capture the digital innovation debate: “Digital Transformation” ({\color{color02}$\sbullet$}) , “Remote Working” ({\color{color03}$\sbullet$})  and “Digitalization” ({\color{color04}$\sbullet$}) . The “Digital Transformation” community deals with the themes of digital transformation, innovation and covid, with the most relevant hashtags being: “smartworking” (“remoteworking”), “cloud”, “covid”, \ale{"ai" and "sicurezza" ("security")}. This community is made of private firms, regionally-owned companies, software companies, consultancies. 
The other two communities (“Remote Working” and “Digitalization”) are mostly concerned with the changing nature of work. One has “covid”, “digitale” (“digital”), “coronavirus”, “lavoro” (“work”), “smartworking” (“remoteworking”) as the most relevant hashtags, which reflect the workplace adjustments during the pandemic. This community is mostly composed of recruitment agencies, along with consultancies, telecommunication companies, a trade fairs organizer, and an agency for the digital innovation. The “Digitalization” community also focuses on work digital adjustments, with “smartworking” (“remoteworking”), “covid”, “lavoro” (“work”), “webinar” and “digitale” (“digital”) as the main hashtags. It is mostly composed of consultancy companies. The “Safety” community ({\color{color08}$\sbullet$}) is made of biopharmaceutical companies, a hospital group and a university consortium. Its hashtags are: “covid”, “italia” (“Italy”), “coronavirus”, “sanità” (“healthcare”), “mascherine” (“facemasks”), thus showing that their main concerns are safety measures against Covid-19.

These results show that the digital innovation debate is central in firms' discussion on Twitter, with themes related to digitalization and the introduction of new tools as remote working, webinars, cloud services. This is consistent with the first results of firms’ reactions to the Covid-19 crisis~\cite{Almeida2020}: the needs to social distancing and to shelter at home pushed towards an acceleration of the digital transformation, which was ongoing before the pandemic, thus changing firms’ business models and strategies. On one hand, this appeared in an increase in the demand for technology products (e.g. laptops) and services (e.g. cloud computers, digital services). On the other hand, many jobs went remote. While digital transformation processes were already ongoing  \cite{matt2015digital}, the pandemic highlighted the need to leverage technology to overcome the challenges and environmental uncertainty, and accelerated a trend towards different life styles, increasing the importance of technology in our economy and society~\cite{Evans2020}.
\fab{Smaller communities are described in details in the appendix~\ref{app:smaller_communities}.}

\subsubsection{CSR dissemination and stakeholder engagement}
{\color{black}
In this second part of the analysis, we study CSR dissemination and stakeholder engagement in the communities previously identified.\\
\fab{Before starting, let us make a few remarks. First, the number of CSR hashtags is extremely limited, being of 30 different words, compared to a set of 6036 different hashtag used by the validated accounts, resulting respectively as the 0.17\% (environmental dimension), 0.28\% (social dimension) and 0.05\% (economic dimension). Despite these small numbers, it is crucial to notice that, when we also consider the repetitions (i.e. the number of times an account used the various hashtags), the fractions are quite different, that is respectively the 1.17\%, 0.42\% and 0.07\%. In this sense, even if the set of keywords we are considering is quite limited, firms' accounts are particularly inclined to use it: if the number of repetitions per hashtag were constant, we would not observe a change in its percentage when considering their total presence. In this sense, also considering the repetitions, the environmental dimension seems particularly popular among the validated accounts.}

\begin{table}[htb]
    \centering
    \begin{tabular}{+c+c|c|c|c|c+}
        \thickhline
        \textbf{Comm.} & \textbf{Accounts} & \textbf{CSR (total)} & \textbf{Env.} & \textbf{Soc.} & \textbf{Econ.}\\
         \thickhline
         {\color{color02} $\sbullet$} & 16 & 9 & 6 & 8 & 2\\
         \hline
         {\color{color01} $\sbullet$} & 14 & 14 & 14 & 8 & 2\\ 
         \hline
         {\color{color03} $\sbullet$} & 11 & 10 & 6 & 9 & 2\\
         \hline
         {\color{color04} $\sbullet$} & 9 & 7 & 5 & 7 & 2\\
         \hline
         {\color{color05} $\sbullet$} & 6 & 4 & 4 & 1 & 0\\
         \hline
         {\color{color08} $\sbullet$} & 6 & 3 & 1 & 2 & 0\\
         \hline
         {\color{color06} $\sbullet$} & 5 & 3 & 3 & 1 & 0\\
         \hline
         {\color{color07} $\sbullet$} & 3 & 3 & 3 & 2 & 2\\
         \hline
         {\color{color11} $\sbullet$} & 2 & 2 & 2 & 0 & 0\\
         \hline
         {\color{color12} $\sbullet$} & 2 & 1 & 1 & 0 & 0\\
         \thickhline
    \end{tabular}
    \caption{\textbf{Frequency of accounts using the various CSR hashtags.} The frequency of accounts using CSR hashtags is higher in the greater communities, with a 100\% covering in the case of ``Environmental Sustainability" ({\color{color01}$\sbullet$}).}
    \label{tab:CSR_fraction}
\end{table}

\paragraph{CSR usage in the validated network}
At the community level, 10 communities out of 13 use CSR hashtags (see Table \ref{tab:CSR_fraction}), with most accounts in each community using CSR hashtags. 
Environmental and social themes are prevalent, with their relevance differing depending on the community. 
In the ``Environmental Sustainability”  community ({\color{color01}$\sbullet$}), all accounts (14) use hashtags related to CSR. This  community is mainly composed of public utilities companies, and highlight that the sustainability debate is prevalent in these kinds of firms, and it is joint with the innovation debate. In this community, all the 14 accounts use hashtags related to the environmental dimension. Hashtags related to the social dimension are also tweeted by most accounts (8 out of 14), while hashtags related to the economic dimension are a minority (tweeted by 2 out of 14 accounts).\\ 
In the ``Digital Transformation” community ({\color{color02}$\sbullet$}), 9 accounts out of 16 use hashtags related to CSR. Among them, 6 discuss environmental themes, 8 social ones and 2 economic ones. This shows that digital transformation themes are more connected to social aspects than environmental ones – among the most frequently used hashtags, words like ``formazione” (``training") and ``istruzione” (``education") appear. Considered together with the prevalent hashtags of the community (``smartworking", ``cloud", ``covid", ...), it seems that firms are discussing the learning process associated with the new technologies, which is consistent with the period of observation.\\
In the ``Remote working” community ({\color{color03}$\sbullet$}), 10 out of 11 accounts use CSR hashtags. Again, the social dimension is prevalent (9 accounts use social hashtags, in contrast to 6 accounts that use environmental hashtags and 2 that use economic hashtags). The hashtags related to the social dimension include training themes (``formazione", ``istruzione”, meaning ``training" and ``education", respectively), but are more varied than the previous community – they concern work-related themes, as ``corruzione”(``corruption"),  ``discriminazione” (``discrimination") and ``diversity”. This is partly intuitive, as this community is mainly composed by recruitment agencies, consultancies, and telecommunication companies.\\ 
In the ``Digitalization" community ({\color{color04}$\sbullet$}), 7 out of 9 accounts use CSR hashtags. Consistently with the previous communities, accounts that use hashtags related to the social dimension are more (7) than the ones posting about the environmental dimension (5), while the economic dimension is still the minor one (2).\\ 
Again, the “Safety” community ({\color{color08}$\sbullet$}) has 2 accounts tweeting about the social dimension and 1 about the environmental dimension, and no accounts tweeting about the economic dimension. 
\newline

These results show that CSR dissemination is not evenly distributed among communities and firms. The prevalent CSR dimension firms disseminate on Twitter is contingent to the community they belong to, which is somehow related to the firms’ sectors – showing that different types of firms emphasize different dimensions of CSR.  
The communities focused on digital innovation (``Digital Transformation", ``Remote Working", ``Digitalization") and safety are more concerned on the social dimension of CSR. In these communities, the environmental dimension is present, but it is less relevant compared to the social dimension, which is somehow surprising, as the environmental dimension is usually the one managers put more attention on~\cite{pedersen2010modelling}. Last, the economic dimension of CSR is overlooked in all communities. 
We do not know if the higher relevance of the social dimension of CSR in the digital innovation and safety communities is an effect of the pandemic or if this also happens in non-crisis times. One option is that the pandemic pushed firms to see the social dimension as the most relevant one, and it became prevalent compared to the environmental one. This would confirm CSR as a concept evolving depending on the present circumstances~\cite{snider2003corporate}. In any case, it is worth noting that, differently from the main literature~\cite{pedersen2010modelling}, in some communities the social dimension is more relevant than the environmental one. Further research could investigate this trend, checking if these communities mostly communicate on social CSR themes over a wider time span, also considering non-crisis times.
\newline

The analysis of the smallest communities concerns for CSR themes can be found in the appendix~\ref{app:smaller_communities_CSR}}

\ale{However, when considering the dissemination of the CSR dimensions focusing on hashtags as a unit of analysis, the relevance of CSR changes. Table~\ref{tab:hash_frequency_by_CSR} shows the occurrence of CSR hashtags among the hashtags in the validated network. Overall, CSR hashtags are less then  2\% of all hashtags, with hashtags related to the environmental dimension being the majority (1.17\%) compared to the social dimension (0.42\%) and the economic one (0.07\%). On one hand, this is consistent with previous literature, which argues that the CSR dimension is overlooked in firms' social media posts \cite{Manetti2016,Gomez2016,Etter2013}. However, a methodological note is needed. As the hashtags we included in the CSR dimensions are only 30 in total, representing the 0.5\% of the total hashtags in the (validated) network, it is easily predictable that they will represent a minority. Actually, their occurrence in the validated network is higher then expected (1.67\%), thus highlighting that firms use the hashtags related to CSR  more frequently then the others. Further research should widen the hashtags considered as related to the CSR dimensions - as for now, we believe that the hashtags we used as representative for the CSR dimension do not entirely capture the phenomenon. 
In any case, some communities show that specific CSR dimensions are more prevalent than others. It is the case of  the "Environmental Sustainability" community ({\color{color01}$\sbullet$}), where hashtags related to the environmental dimension are the 3.51\% of the total. Firms in the "{Remote Working}" community ({\color{color03}$\sbullet$}), instead, use more hashtags related to the social dimension, which account for the 1.12\% of the total. Again, the "{Restart}" community ({\color{color04}$\sbullet$}) shows a higher prevalence of the environmental dimension (1.62\%). }

\begin{table}[htb!]
    \centering
    \begin{tabular}{+c|c|c|c|c+}
    \thickhline
        & \textbf{Hashtags} & \textbf{ENV (\%)} & \textbf{SOC (\%)} &\textbf{ECON (\%)}\\
        \thickhline
         Validated network & 67805 & 1.17 & 0.42 & 0.07\\
         {\color{color02} $\sbullet$} & 12646 & 0.38 &0.29 & 0.05\\
         {\color{color01} $\sbullet$} & 14541 & 3.51 &0.50 & 0.04\\
         {\color{color03} $\sbullet$} & 10921 & 0.51 &1.12 & 0.07\\
         {\color{color04} $\sbullet$} & 7164 & 1.62 &0.52 & 0.08\\
         {\color{color05} $\sbullet$} & 3312 & 0.63 &0.06 & 0.00\\
         {\color{color08} $\sbullet$} & 4751 & 0.04 &0.08 & 0.00\\
         {\color{color06} $\sbullet$} & 8223 & 0.32 &0.02 & 0.00\\
         {\color{color07} $\sbullet$} & 2115 & 0.28 &0.47 & 1.13\\
         {\color{color00} $\sbullet$} & 1142 & 0.00 &0.00 & 0.00\\
         {\color{color09} $\sbullet$} & 272 & 0.00 &0.00 & 0.00\\
         {\color{color10} $\sbullet$} & 10 & 0.00 &0.00 & 0.00\\
         {\color{color11} $\sbullet$} & 1708 & 0.35 &0.00 & 0.00\\
         {\color{color12} $\sbullet$} & 1000 & 0.40 &0.00 & 0.00\\
         \thickhline
    \end{tabular}
    \caption{\fab{\textbf{Frequency of hashtags in the validated network, by community and CSR dimension}. As expected by analysing the community composition in terms of ATECO code, the attention on the various group is quite different: for instance, the ``Environmental Sustainability" community {\color{color01} $\sbullet$} has a frequency of hashtags in the Environmental dimension that is greater than twice the value observed on the entire validated network. Similar considerations apply to the ``Remote Working" community {\color{color03} $\sbullet$} for the social dimension and the ``Mobility" one {\color{color07} $\sbullet$} for the social one.}}
    \label{tab:hash_frequency_by_CSR}
\end{table}

\begin{table}[htb!]
    \hskip-2cm
    \begin{tabular}{+c|c|c|c+c|c|c|c+}
    \thickhline
    \multicolumn{4}{+c+}{\textbf{Likes}} & \multicolumn{4}{c+}{\textbf{Retweets}} \\
    \thickhline
    & \textbf{ENV} & \textbf{SOC} & \textbf{ECON} & & \textbf{ENV} & \textbf{SOC} & \textbf{ECON}\\
    \thickhline
         Validated network & 2.42 & 2.76 &1.47 & Validated network & 2.44 & 3.69 &1.79\\
         \thickhline
         {\color{color02} $\sbullet$} & 2.75 & 2.22 &0.80 & {\color{color02} $\sbullet$} & 2.49 & 2.88 &1.70\\
         \hline
         {\color{color01} $\sbullet$} & 2.27 & 3.13 &1.79 & {\color{color01} $\sbullet$} & 2.31 & 3.80 &2.03\\
         \hline
         {\color{color03} $\sbullet$} & 2.20 & 2.84 &0.80 & {\color{color03} $\sbullet$} & 2.33 & 3.99 &1.70\\
         \hline
         {\color{color04} $\sbullet$} & 2.50 & 2.72 &1.27 & {\color{color04} $\sbullet$} & 2.51 & 3.91 &1.39\\
         \hline
         {\color{color05} $\sbullet$} & 2.72 & 1.31 &0.00 & {\color{color05} $\sbullet$} & 2.79 & 1.89 &0.00\\
         \hline
         {\color{color08} $\sbullet$} & 2.83 & 2.39 &0.00 & {\color{color08} $\sbullet$} & 2.92 & 2.85 &0.00\\
         \hline
         {\color{color06} $\sbullet$} & 2.55 & 5.03 &0.00 & {\color{color06} $\sbullet$} & 2.46 & 4.86 &0.00\\
         \hline
         {\color{color07} $\sbullet$} & 2.83 & 2.81 &2.78 & {\color{color07} $\sbullet$} & 2.92 & 4.02 &2.35\\
         \hline
         {\color{color00} $\sbullet$} & 0.00 & 0.00 &0.00 & {\color{color00} $\sbullet$} & 0.00 & 0.00 &0.00\\
         \hline
         {\color{color09} $\sbullet$} & 0.00 & 0.00 &0.00 & {\color{color09} $\sbullet$} & 0.00 & 0.00 &0.00\\
         \hline
         {\color{color10} $\sbullet$} & 0.00 & 0.00 &0.00 & {\color{color10} $\sbullet$} & 0.00 & 0.00 &0.00\\
         \hline
         {\color{color11} $\sbullet$} & 2.83 & 0.00 &0.00 & {\color{color11} $\sbullet$} & 2.92 & 0.00 &0.00\\
         \hline
         {\color{color12} $\sbullet$} & 2.88 & 0.00 &0.00 & {\color{color12} $\sbullet$} & 3.23 & 0.00 &0.00\\
         \thickhline
    \end{tabular}
    \caption{\fab{\textbf{Average number of likes (left) and retweets (right) per message containing a hashtag in the environmental (\textbf{ENV}), social (\textbf{SOC}) and economic (\textbf{ECON}) dimensions.} Interestingly enough, the frequency of the CSR hashtags used (reported in Table~\ref{tab:hash_frequency_by_CSR}) is not necessarily mirrored in the number of likes or retweets received on average per hashtag.}}
    \label{tab:user_engagement}
\end{table}

\paragraph{Stakeholder engagement on CSR dimensions}

\ale{We measure stakeholder (user) engagement as the number of retweets and likes per hashtag. 
Table~\ref{tab:user_engagement} shows stakeholder engagement on all posts and CSR dimensions. Results show that posts with hashtags related to the three CSR dimensions are retweeted and liked less then the rest of the posts, thus confirming that social media are not fully exploited to interact and engage on CSR themes \cite{Gomez2016,Manetti2016,Etter2013}.


In the entire data set, on average, each message containing a hashtag concerning CSR is retweeted 2.99 times, against an average of 5.39 retweets for all the hashtags. However, a closer look at the communities and CSR dimensions highlights some peculiarities. While the number of retweets for the environmental dimension is more or less stable in the communities, the ``Mobility" ({\color{color06}$\sbullet$}) community has a higher number of retweets in the social dimension (4.86 on average). With a few exceptions, the social dimension has on average a higher number of retweets compared to the other CSR dimensions and to the dataset.\\ 
In terms of the number of likes per hashtag, in the entire dataset each message containing a CSR hashtag obtains on average  2.59 likes, against an average of 14.83 likes for all the hashtags. When we look at the various communities, again the number of likes in the environmental dimension is quite stable, while social hashtags show a few different trends, with the ``Mobility" ({\color{color06}$\sbullet$}) and ``Environmental Sustainability" ({\color{color01}$\sbullet$}) communities having a higher number of likes (5.03 and 3.13 respectively) and the ``Restart" one ({\color{color05}$\sbullet$}) having a lower number of likes than the average (1.31).  
Thus, it seems that stakeholder engagement shows similar trends with both measures (retweets and likes), while the extent to which users interact is unevenly distributed among dimensions and communities.} 
\newline

\ale{Overall, these findings show that firms' dissemination on CSR dimensions and stakeholder engagement are not homogeneous and vary depending on the communities. Thus, studies enquiring on firms' CSR concerns and stakeholder engagement should consider this when formulating their research questions and methods. These findings also show that network methods, allowing firms' discussions to emerge from data without any a priori hypothesis are an effective way to show firms' and users' different attitudes towards the CSR dimensions. }

\section{Conclusion}

Our paper presents large Italian firms' 
discussion on Twitter during the first 9 months of the Covid-19 pandemic. Specifically, our explorative research questions followed three lines: 1) firms’ general Twitter discussion during the Covid-19 pandemic; 2) the CSR dimensions being discussed; 3) stakeholder engagement on CSR themes. 
First, we show that the discussion is formed of 13 communities of firms, with Covid-19 themes appearing in 10 of them. The core of the network, which reflects firms’ major challenges during the pandemic, is composed of five communities (i.e. ``Digital Transformation", ``Remote Working", ``Digitalization", ``Environmental Sustainability" and ``Safety"). Firms' ownership type does not seem to affect the debate. Second, we show that 10 communities out of 13 use CSR hashtags. While the environmental and the social dimensions are the prevalent ones, the economic dimension is generally overlooked. Moreover, the social dimension is more relevant than the environmental one in the three digital innovation communities and the safety one. Third, stakeholder engagement on CSR themes is limited, and unevenly distributed among communities and themes.

This work has methodological, theoretical and practical implications.
On the methodological side, we integrate a new methodology (i.e. complex network analysis) in management research, widening the methods and focus of current research, that mostly uses manual labelling \cite{Manetti2016}, and linear regressions \cite{Yang2020,Gomez-Carrasco2020,Yang2017} and centers on specific \cite{Pilar2019,sharma2020covid} or CSR themes~\cite{Chae2018a,Colleoni2013,Pilar2019}). 

\alee{Theoretically, we contribute to the literature on firms and social media, showing that digital innovation, environmental sustainability and safety are the main challenges firms are facing during Covid-19; to the CSR literature \cite{Manetti2016,giacomini2020environmental}, highlighting that the relevance of the CSR dimensions varies depending on the community; and to the stakeholder engagement literature \cite{manetti2017stakeholder,bonson2016corporate,okazaki2020exploring}, confirming that Twitter is overlooked as a tool to interact on CSR issues, with peculiarities arising in some communities.
\par

Businesses today use social media as predictive tools on various sides, including product design, relations management and marketing \cite{lu2014integrating}. For management scholars, social media are a promising tool for a series of aspects, including stock market volatility \cite{antweiler2004all}, product sales \cite{ghose2010estimating}, and financial performances \cite{kim2017social}. 
However, the potential of social media for strategic decision making is widely overlooked. Today, social media can provide businesses a rich source of naturally occurring data \cite{ruhi2014social}, which can be exploited to detect emerging issues and relevant themes.
Thus, on the practical side, we believe our research contributes to provide an analytics tool based on firms’ Twitter data to support managers, entrepreneurs and policy makers when designing their strategies and decision making.}
Applied to the CSR field, a network-based analysis of Twitter represents an alternative way to understand the managerial perceptions of CSR themes, which are the responsibilities managers believe a business should pursue towards society~\cite{pedersen2010modelling}.

Some further questions remain to be addressed. First, our analyses focus on large firms in only one country, thus not considering small and medium businesses, which form the backbone of Italian and European economies, and other national contexts. An extension of this research will also include SMEs, to allow a wider understanding of firm's discussion on Twitter.  In fact, due to the selection of large firms, firms' sectors are not evenly distributed, neither is firms' geographical location. A larger set of firms' account will overcome this limitation.\\

Second, the examined list of CSR hashtags is quite limited and, even if those hashtags are quite popular, we run the risk of capturing just a portion of the entire CSR debate; in further research we will expand this list, increasing our coverage of all the variations of the CSR discussion.

\section*{Acknowledgments}
GC acknowledges from the EU grant ``Humane-AI-net" (Grant nr. 952026). \fab{FS acknowledges support from the EU project SoBigData++ (Grant nr. 871042) and the PAI (\emph{Progetto di Attivit\`a Integrata}) project TOFFEe, funded by the IMT School For Advanced Studies Lucca}.

\nolinenumbers

\appendix

\section{Hashtag cleaning}\label{app:hashtags}
In order to overcome typos and to recognise properly the same subjects expressed by different hashtags, we implemented the edit distance, as said in the main text. Before implementing the edit distance, we first detect words containing numbers, which may represent dates or anniversaries: in this sense, we delete numbers and compare the non-numeric characters of the hashtags, in order to focus on the relevant information. The rationale is that we are interested in focusing on the fact that for instance, a firm is celebrating its anniversary to stress its robustness, but not on how many years they are celebrating. We do not consider in the comparison possible acronyms, i.e. words that were not recognise to be Italian or English, i.e. the two main languages for the company communication in Italy: it was done in order to avoid considering the random match of different acronyms.

Hashtags with a relative edit distance smaller than 0.20 were considered equal. The value of 0.20 was chosen after manually testing a sample of 200 couple of hashtags with edit distance lower than 0.20: the 90.9\% of identifications were correct. We tried other thresholds, but the performances were worse. The identified hashtags were subsequently merged, after a manual check. 
\newline

\section{Entropy-based null-models for bipartite networks}\label{app:entropy}
\fab{In the present section of the appendix we present more details about the  the entropy-based null-model used for the analysis of the bipartite network between the layer of firm accounts and the one of (cleaned) hashtags, i.e. the Bipartite Configuration Model (\emph{BiCM}~\cite{Saracco2015}) mentioned in Section~\ref{ssec:bicm}; a general review on the wider argument of entropy-based null-models for the analysis of real networks can be found in~\cite{Cimini2018}. After the introduction of the BiCM in  subsection~\ref{app:bicm}, we will show in subsection~\ref{app:inferring} how the null-model can be used as a benchmark to validate the projection on one of the two layers, as proposed in~\cite{Saracco2017}. Let us finally remark that the exact implementation of the null-model was performed via the python module {\tt\href{https://pypi.org/project/NEMtropy/}{NEMtropy}}, presented in~\cite{vallarano2021fast}.
}

\fab{
\subsection{Bipartite Configuration Model}\label{app:bicm}
Let us start from a real bipartite network $\mathbf{G}^*_\text{Bi}$ and call the two layers $\top$ and $\bot$; we refer to nodes of the layer $\top$ and $\bot$ using respectively with Latin and Greek indices. A bipartite network can be represented via its biadjacency matrix, i.e. a rectangular $|\top|\times|\bot|$-matrix $\mathbf{M}$ whose the generic entry $m_{i\alpha}$  is 1 if $i\in\top$ and $\alpha\in\bot$ are connected and zero otherwise. The degree of nodes $i$ and $\alpha$ can be written in terms of the biadjacency matrix as $k_i=\sum_{\alpha\in\bot}m_{i\alpha}$ and $k_\alpha=\sum_{i\in\top}m_{i\alpha}$, respectively.\\
The main idea is to build a network null-model that has some topological properties (in the case of the BiCM, the degree sequence) equal, on average, to the ones observed in the real system and is completely random for the remaining ones, as the statistical ensembles in Statistical Mechanics~\cite{Cimini2018}. Thus, let us call $\mathcal{G}_\text{Bi}$ the set (the \emph{ensemble}) of all bipartite graphs with the same number of nodes, respectively in $\top$ and $\bot$, as in the real network $\mathbf{G}^*_\text{Bi}$ and let us define the Shannon entropy for the ensemble~\cite{park2004statistical}:
\begin{equation}\label{eq:S}
S=-\sum_{\textbf{G}_\text{Bi}\in\mathcal{G}_\text{Bi}}P(\textbf{G}_\text{Bi})\ln P(\textbf{G}_\text{Bi}), 
\end{equation}
where $P(\textbf{G}_\text{Bi})$ is the probability of the representative $\textbf{G}_\text{Bi}\in\mathcal{G}_\text{Bi}$. In order to consider the null-model as maximally random, but for the constraints, we have to maximise the entropy in Eq.~\ref{eq:S} constraining the average value of the degree sequence. We can obtain this result by using Lagrangian multipliers, i.e. by finding the maximum of 
\begin{equation*}
    S'=S+\zeta\Big(1-\sum_{\textbf{G}_\text{Bi}\in\mathcal{G}_\text{Bi}}P(\textbf{G}_\text{Bi})\Big)+\sum_i \eta_i \Big(k_i^*-\langle k_i\rangle\Big)+\sum_\alpha \theta_\alpha \Big(h_\alpha^*-\langle h_\alpha\rangle\Big),
\end{equation*}
where $\langle...\rangle$ are averages over the ensemble, $\zeta$, $\eta_i$ and $\theta_\alpha$ are the Lagrangian multipliers and $\ast$ indicates values measured over the real network. The maximisation of the entropy respect to the probability returns the functional forms of the probability in terms of the Lagrangian multipliers $\zeta$, $\eta_i$ and $\theta_\alpha$. Interestingly enough, when the constraints are linear in the adjacency matrix (as in the case of the degree sequence), the probability for the entire graph factorises in terms of probability per links, i.e. 
\begin{equation*}
P(\mathbf{G}_\text{Bi}) = \prod_{i,\alpha} \left(p_{i\alpha}\right)^{m_{i\alpha}}\left(1-p_{i\alpha}\right)^{1-m_{i\alpha}},
\end{equation*}
where $p_{i\alpha}$ is the probability of observing a link in $m_{i\alpha}$.

The Lagrangian multiplier $\zeta$ is simply the normalisation of the probability per graph $P(\mathbf{G}_\text{Bi})$. Instead, to finally obtain the value of $\eta_i$ and $\theta_\alpha$, we have to impose that the average degree sequence is equal to the one observed in the real system, i.e. 
\begin{equation}\label{eq:likelihood}
\left\{
\begin{split}
\left\langle k_i \right\rangle &= k_i^* \quad  \forall i \in \top\\
\left\langle k_\alpha \right\rangle &=k_\alpha^* \quad \forall  \alpha \in \bot.\\
\end{split}
\right..
\end{equation}
Interestingly enough, it has been shown that the system of Eq.~\ref{eq:likelihood} is analogous to the maximisation of the likelihood of the observed system~\cite{Garlaschelli2008,squartini2011analytical}.}

\begin{table}[htb!]
    \hskip-5.0cm
    \begin{tabular}{+c|c|m{4cm}|c+c|c|m{4cm}|c+}
        \thickhline
        \textbf{Comm.} & \textbf {ATECO} & \textbf{Description} & \textbf{Occ.} & \textbf{Comm.} & \textbf {ATECO} & \textbf{Description} & \textbf{Occ.}\\
        \thickhline
        {\color{color02} $\sbullet$} & 62 & computer programming, consultancy and related activities & 7& {\color{color03} $\sbullet$} & 78 & employment activities & 3\\ \cline{2-4} \cline{6-8} & 26 & manufacture of computer, electronic and optical products & 3& & 62 & computer programming, consultancy and related activities & 2\\ \cline{2-4} \cline{6-8} & 46 & wholesale trade, except of motor vehicles and motorcycles & 2& & 61 & telecommunications & 1\\ \cline{2-4} \cline{6-8} & 43 & specialised construction activities & 1& & 63 & information service activities & 1\\ \cline{2-4} \cline{6-8} & 52 & warehousing and support activities for transportation & 1& & 70 & activities of head offices; management consultancy activities & 1\\ \cline{2-4} \cline{6-8} & 82 & office administrative, office support and other business support activities & 1& & 82 & office administrative, office support and other business support activities & 1\\ \cline{2-4} \cline{6-8} & 91 & libraries, archives, museums and other cultural activities & 1& & 84 & public administration and defence; compulsory social security & 1\\ \cline{1-4} \cline{6-8}{\color{color01} $\sbullet$} & 35 & electricity, gas, steam and air conditioning supply & 5& & 85 & education & 1\\ \cline{2-4} \cline{5-8} & 71 & architectural and engineering activities; technical testing and analysis & 3& {\color{color04} $\sbullet$} & 70 & activities of head offices; management consultancy activities & 3\\ \cline{2-4} \cline{6-8} & 27 & manufacture of electrical equipment & 2& & 63 & information service activities & 2\\ \cline{2-4} \cline{6-8} & 20 & manufacture of chemicals and chemical products & 1& & 78 & employment activities & 2\\ \cline{2-4} \cline{6-8} & 36 & water collection, treatment and supply & 1& & 69 & legal and accounting activities & 1\\ \cline{2-4} \cline{6-8} & 46 & wholesale trade, except of motor vehicles and motorcycles & 1& & 73 & advertising and market research & 1\\ \cline{2-4} \cline{6-8} & 49 & land transport and transport via pipelines & 1 &&&&\\
        \thickhline
        \end{tabular}
    \caption{\textbf{ATECO codes of the accounts involved in the various communities in the validated network of Fig.~\ref{fig:network}, 1/2.}}
    \label{tab:ATECO_0}
\end{table}

\subsection{Validated projection}\label{app:inferring}
\fab{We can use the BiCM to study the similarity of the connections of the nodes on one of the layers and detect the significant similarities. 
Consider two nodes $i,j\in\top$: such similarity can be captured by the co-occurrences of links towards the opposite layer, or \emph{V-motifs}~\cite{Saracco2015}:
\begin{equation*}
    V_{ij} = \sum_{\alpha} m_{i\alpha}m_{j\alpha}. 
\end{equation*}
The method presented in~\cite{Saracco2017} proposes to compare, for each couple of nodes $i,j\in\top$, the observed V-motifs with the theoretical distribution obtained using the BiCM probabilities. The distribution of V-motifs is a Poisson binomial one, i.e. the generalization of a binomial distribution in which each event has a different probability; in the present application, due to the sparsity of our network, it can be safely approximate the Poisson binomial distribution with a Poisson one~\cite{Hong2013}.\\
Once we have the theoretical distribution of the V-motifs we can state the statistical significance of their observation: for each couple $i,j\in\top$, we can associate a p-value related to the observed $V_{ij}^*$. We finally need a multiple test hypothesis to state the statistical significance of the p-values: the False Discovery Rate (\emph{FDR}) procedure~\cite{benjamini1995controlling} is considered as one of the most effective, due to its control on the False Positive rate. All rejected hypotheses by the FDR indicate that the relative V-motifs cannot be explained by the theoretical distribution, that is, they contain more information than those of the constraints. Following this rationale, we can define a validated projection network as a monopartite network for nodes on layer $\top$ where a link between $i$ and $j$ is present if the related $V_{ij}^*$ is statistically significant.
}

\begin{table}[htb!]
    \hskip-5.0cm
    \begin{tabular}{+c|c|m{4cm}|c+c|c|m{4cm}|c+}
        \thickhline
        \textbf{Comm.} & \textbf {ATECO} & \textbf{Description} & \textbf{Occ.} & \textbf{Comm.} & \textbf{ATECO} & \textbf{Description} & \textbf{Occ.}\\
        \thickhline
        {\color{color05} $\sbullet$} & 47 & retail trade, except of motor vehicles and motorcycles & 3& {\color{color07} $\sbullet$} & 21 & manufacture of basic pharmaceutical products and pharmaceutical preparations & 3\\ \cline{2-4}\cline{5-8} & 79 & travel agency, tour operator and other reservation service and related activities & 2& {\color{color00} $\sbullet$} & 88 & social work activities without accommodation & 2\\ \cline{2-4}\cline{5-8} & 51 & air transport & 1& {\color{color09} $\sbullet$} & 61 & telecommunications & 1\\ \cline{1-4}\cline{6-8}{\color{color08} $\sbullet$} & 86 & human health activities & 3& & 71 & architectural and engineering activities; technical testing and analysis & 1\\ \cline{2-4}\cline{5-8} & 21 & manufacture of basic pharmaceutical products and pharmaceutical preparations & 2& {\color{color10} $\sbullet$} & 10 & manufacture of food products & 1\\ \cline{2-4}\cline{6-8} & 62 & computer programming, consultancy and related activities & 1& & 78 & employment activities & 1\\ \cline{1-4}\cline{5-8}{\color{color06} $\sbullet$} & 52 & warehousing and support activities for transportation & 2& {\color{color11} $\sbullet$} & 50 & water transport & 1\\ \cline{2-4}\cline{6-8} & 42 & civil engineering & 1& & 52 & warehousing and support activities for transportation & 1\\ \cline{2-4}\cline{5-8} & 50 & water transport & 1& {\color{color12} $\sbullet$} & 47 & retail trade, except of motor vehicles and motorcycles & 1\\ \cline{2-4}\cline{6-8} & 70 & activities of head offices; management consultancy activities & 1& & 58 & publishing activities & 1\\ 
        \thickhline
        \end{tabular}
    \caption{\textbf{ATECO codes of the accounts involved in the various communities in the validated network of Fig.~\ref{fig:network}, 2/2.}}
    \label{tab:ATECO_1}
\end{table}

\section{Smaller communities}\label{app:smaller_communities}

\begin{table}[htb!]
\hskip-4.0cm    
    \begin{tabular}{+c|c|c+c|c|c+}
        \thickhline
        \textbf{Comm.} & \textbf{Hashtag} & \textbf{Occurrence}& \textbf{Comm.} & \textbf{Hashtag} & \textbf{Occurrence}\\
        \thickhline 
          {\color{color07}  $\sbullet$} & iorestoacasa & 3 & {\color{color11}  $\sbullet$} & italia & 2 \\ 
          \cline{2-3} \cline{5-6} 
          & covid & 3 & & brindisi & 2 \\ 
          \cline{2-3} \cline{5-6} 
          & coronavirus & 3 & &  sostenibilità & 2 \\ 
          \cline{2-3}\cline{5-6} 
          & pandemia & 3 & & bari & 2 \\ 
          \cline{2-3}\cline{5-6} 
          & psoriasi & 3 & & cagliari & 2 \\ 
          \cline{2-3}\cline{5-6} 
          & sostenibilità & 3 & & sicilia & 2 \\ 
          \cline{2-3}\cline{5-6} 
          & covid19italia & 3 & & napoli & 2 \\ 
          \cline{2-3}\cline{5-6} 
          & noncifermeremo & 3 &  & genova & 2 \\ 
          \cline{1-3}\cline{5-6} 
          {\color{color00}  $\sbullet$} & maggio & 2 & & trieste & 2 \\ 
          \cline{2-3} \cline{5-6} 
          & fratellitutti & 2 &  & catania & 2\\ 
          \cline{2-3}\cline{5-6} 
          & marzo & 2 &  & palermo & 2 \\ 
          \cline{2-3} \cline{5-6} 
          & servizi & 2 &  & coronavirus & 2\\ 
          \cline{2-3} \cline{5-6} 
          & settembre & 2 &  & covid & 2\\ 
          \cline{2-3} \cline{4-6} 
          & luglio & 2 & {\color{color12}  $\sbullet$} & nuda & 2 \\ 
          \cline{2-3} \cline{5-6} 
          & buonapasqua & 2 & & nitro & 2 \\ 
          \cline{2-3} \cline{5-6} 
          & giornatamondialedeipoveri & 2 & & primavera & 2 \\ 
          \cline{2-3} \cline{5-6} 
          & livatino & 2 &  & scuola & 2 \\ 
          \cline{2-3} \cline{5-6} 
          & giugno & 2 &  & naked & 2 \\ 
          \cline{2-3} \cline{5-6} 
          & libano & 2 &  & maggiodeilibri2020 & 2 \\ 
          \cline{2-3} \cline{5-6} 
          & ottobre & 2 &  & maggio & 2 \\ 
          \cline{2-3} \cline{5-6} 
          & coronavirus & 2 &  & ozpetek & 2 \\ 
          \cline{2-3} \cline{5-6} 
          & covid & 2 &  & backtoschool & 2 \\ 
          \cline{2-3} \cline{5-6} 
          & iorestoacasa & 2 &  & unestateinnero & 2 \\ 
          \cline{2-3} \cline{5-6} 
          & ansa & 2 &  & nek & 2 \\ 
          \cline{2-3} \cline{5-6} 
          & pandemia & 2 &  & michelebravi & 2 \\ 
          \cline{2-3} \cline{5-6} 
          & papafrancesco & 2 &  & thriller & 2 \\ 
          \cline{2-3} \cline{5-6} 
          & aprile & 2 &  & chatroom & 2\\ 
          \cline{1-3}\cline{5-6} 
          {\color{color09}  $\sbullet$} & csg & 2 &  & giugno & 2 \\ 
          \cline{2-3} \cline{5-6} 
          & e\_geos & 2 & & aliciakeys & 2 \\ 
          \cline{2-3} \cline{5-6} 
          & covid & 2 &  & iorestoacasa & 2 \\ 
          \cline{2-3} \cline{5-6} 
          & cosmoskymed & 2 &  & libridaleggere & 2 \\ 
          \cline{2-3} \cline{5-6} 
          & copernicus & 2 &  & lercio & 2 \\ 
          \cline{2-3} \cline{5-6} 
          & comunicatostampa & 2 & & libri & 2\\ 
          \cline{2-3} \cline{5-6} 
          & climatechange & 2 &  & estate & 2\\ 
          \cline{2-3}\cline{5-6} 
          & space & 2 &  & kenfollett & 2 \\ 
          \cline{2-3} \cline{5-6} 
          & spaceeconomy & 2 & & giornatamondialedellibro & 2 \\ 
          \cline{2-3} \cline{5-6} 
          & sky & 2 &  & yoga & 2 \\ 
          \cline{2-3} \cline{5-6} 
          & spazio & 2 &  & raffaellosanzio & 2\\ 
          \cline{2-3}\cline{5-6} 
          & telespazio & 2 &  & forestami & 2 \\ 
          \cline{2-3}\cline{5-6} 
          & terra & 2 &  & nakedroom & 2 \\ 
          \cline{2-3}\cline{5-6} 
          & leonardo & 2 &&&\\ 
          \cline{2-3}\cline{5-6} 
          & ambiente & 2 &&&\\ 
          \cline{1-3}\cline{5-6} 
          {\color{color10}  $\sbullet$} & coronavirus & 2 &&&\\ 
          \cline{2-3} \cline{5-6} 
          & covid & 2 &&&\\ 
          \cline{2-3} \cline{5-6} 
          & rsa & 1 &&&\\ 
         \thickhline
    \end{tabular}
    \caption{{\bf The top 5 frequent hashtag for the smaller communities of the network of Fig.~\ref{fig:network}}. Due to the great number of \emph{ex aequo}, the top 5 most frequent hashtags is, in general, longer than 5. }
    \label{tab:hashtag_1}
\end{table}

The smaller communities in the validated network either deal with Covid-19 themes or topics specific to firms’ sectors, see Table~\ref{tab:hashtag_1}. The “Restart” community ({\color{color05}$\sbullet$}) has Italy and coronavirus as its main themes, with “ripartiamodallitalia” (“restartfromitaly"), “covid”, “italia” (“Italy”), “coronavirus”, “estatepostcovid” (“summeraftercovid”) as the most frequent hashtags. This community is composed of a tour operator, an airline and supermarkets. In the “Mobility” community ({\color{color06}$\sbullet$}), the main hashtags are “settembre” (“September”), “agosto” (“August”), “ottobre” (“October”), “maggio” (“May”), “marzo” (“March”). It is made of firms active in the mobility sector: mainly railways, highways, roads managers and other transport means. They mostly use Twitter to share news, and their hashtags are the months when the communication takes place, consistently with what previous research found \cite{manetti2017stakeholder}. In the community “Slowing the Spread” ({\color{color07}$\sbullet$}), the main hashtags are: “iostoacasa” (“stayhome”), “coronavirus”, “psoriasis”, “sostenibilità” (“sustainability”), “noncifermeremo” (“wewillnotstop”). Thus, it is mainly focused on Covid-19 related themes, illnesses and sustainability. Consistently with the themes, it is composed by biopharmaceutical companies.  “Space” ({\color{color09}$\sbullet$}) is the first of the micro-communities, which focuses on space and the environment. It is composed of two firms, one dealing with geoinformation services and applications, the other with spaceflight services. Their hashtags are: “cosmoskymed”, “climatechange”, “space”, “Telespazio”, “ambiente” (“environment”), thus showing that the Covid-19 pandemic has not entered the debate. 
Another small community (“Nursing Homes”, {\color{color10}$\sbullet$}) is formed by two different companies (a producer of canned tuna and a caregiver agency). Their most used hashtags are: “coronavirus”, “covid”, “rsa” (“nursinghome”), which mainly reflect the caregiver agency themes. The community “Locality” ({\color{color11}$\sbullet$}) is formed by two companies in the leisure travel segment (a cruise line, an airport). Its hashtags reflect the places where they are settled: “Italia” (“Italy”), “sostenibilità” (“sustainability”), “Sicilia” (“Sicily”), “Catania”, “Palermo”. The “Reading” community ({\color{color12}$\sbullet$}) is made of two book companies. Its hashtags are: “maggiodeilibri2020” (“Maybooks2020), “libridaleggere” (“bookstoread”), “iorestoacasa” (“stayhome”), “unestateinnero” (“asummerinblack”), “backtoschool”.  The last community is “Solidarity” ({\color{color00}$\sbullet$}), as the discussion is based on solidarity and Covid-19 themes, with “coronavirus”, “fratellitutti” (“allbrothers”), “giornatamondialedeipoveri” (“worldpovertyday”), “iorestoacasa” (“stayhome”), “papafrancesco” (“popefrancis”) as the main hashtags.

\section{CSR usage in the smaller communities}\label{app:smaller_communities_CSR}
\ale{The remaining communities – which are smaller – all show some interest in CSR themes. The economic dimension of CSR is absent, except in the “Mobility” community, where two accounts use the hashtag “responsabilità” ("responsibility"). In all the minor communities the environmental dimension is bigger than the social one, with “sostenibilità” ("sustainability") as the recurring hashtag. 
As far as specific hashtags are concerned, these are varied depending on the community (see Table~\ref{tab:CSR_key}). “Sostenibilità” (“sustainability”) is the more common hashtag in the “environmental” area, which appears in all the 10 communities.  
The two biggest communities (“Digital Transformation” and “Environmental Sustainability”) include “stakeholder” as their most frequently tweeted hashtags. This means that firms are specifically addressing their stakeholders~\cite{freeman2015stakeholder}.
 “Csr” e “responsabilità” (“responsibility”) are the only hashtags used in the "economic” area, with the latter one appearing in only three communities. As the concept of responsibility is connected to the idea of accountability, the low presence of this hashtag shows that Twitter is not used as a mean to report firm’s economic and financial results. This is consistent with previous research, which maintains that firms use Twitter mostly for marketing purposes and to promote their products and service, rather than for the sharing news about CSR \cite{Gomez2016}.
}

\begin{table}[htb!]
    \centering
    \begin{tabular}{+c+m{3cm}| m{3cm}|m{3cm}+}
        \thickhline
        \textbf{Comm.} &  \textbf{Env.} & \textbf{Soc.} & \textbf{Econ}\\
        \thickhline
        {\color{color02} $\sbullet$} & energia, energy, rifiuti, sostenibilità, sustainability & compliance, comunità, engagement, formazione, istruzione, stakeholder, training & csr\\ \hline {\color{color01} $\sbullet$} & acqua, biodiversità, emissioni, energia, energy, rifiuti, sostenibilità, sustainability & coinvolgimento, diversity, education, formazione, fornitori, istruzione, occupazione, stakeholder & csr, responsabilità\\ \hline {\color{color03} $\sbullet$} & acqua, biodiversità, emissioni, energia, energy, rifiuti, sostenibilità & corruzione, dipendenti, discriminazione, diversity, formazione, istruzione, occupazione & csr\\ \hline {\color{color04} $\sbullet$} & energia, energy, rifiuti, sostenibilità & community, discriminazione, diversity, education, engagement, etica, formazione, fornitori, istruzione, occupazione & csr, responsabilità, tax\\ \hline {\color{color05} $\sbullet$} & rifiuti, sostenibilità & comunità & \\ \hline {\color{color08} $\sbullet$} & sostenibilità & formazione, training & \\ \hline {\color{color06} $\sbullet$} & energia, rifiuti, sostenibilità & diversity & \\ \hline {\color{color07} $\sbullet$} & sostenibilità & community, diversity, formazione, occupazione & responsabilità\\ \hline {\color{color11} $\sbullet$} & sostenibilità &  & \\ \hline {\color{color12} $\sbullet$} & biodiversità, sostenibilità &  & \\
        \thickhline

    \end{tabular}
    \caption{\textbf{CSR keywords per community.} As expected, the various CSR dimensions used depends on the ATECO sectors covered by the community.}
    \label{tab:CSR_key}
\end{table}

\end{document}